\documentclass[%
    reprint,
    superscriptaddress,
    showpacs,
    amsmath,
    amssymb,
    aps,
    prd,
]{revtex4-2}

\usepackage{graphicx}
\usepackage{wrapfig}
\usepackage[caption=false]{subfig}
\usepackage{xcolor} 
\usepackage{amssymb} 
\usepackage{hyperref} 
\usepackage[utf8]{inputenc} 
\usepackage[english]{babel} 
\usepackage{blindtext} 
\usepackage{subfiles}
\usepackage{multirow}
\usepackage{booktabs}
\usepackage{bm}
\usepackage{cases}
\usepackage{autobreak}
\usepackage{tabularx}
\usepackage{comment}
\usepackage{orcidlink} 

\graphicspath{{Figures/}} 

\makeatletter
\renewcommand{\p@subsection}{}
\renewcommand{\p@subsubsection}{}
\makeatother

\begin{document}

\preprint{APS/123-QED}

\title{A test of lepton flavor universality with a measurement of $R(D^{*})$ using hadronic $B$ tagging at the Belle~II experiment}
  \author{I.~Adachi\,\orcidlink{0000-0003-2287-0173}} 
  \author{K.~Adamczyk\,\orcidlink{0000-0001-6208-0876}} 
  \author{L.~Aggarwal\,\orcidlink{0000-0002-0909-7537}} 
  \author{H.~Ahmed\,\orcidlink{0000-0003-3976-7498}} 
  \author{H.~Aihara\,\orcidlink{0000-0002-1907-5964}} 
  \author{N.~Akopov\,\orcidlink{0000-0002-4425-2096}} 
  \author{A.~Aloisio\,\orcidlink{0000-0002-3883-6693}} 
  \author{N.~Anh~Ky\,\orcidlink{0000-0003-0471-197X}} 
  \author{D.~M.~Asner\,\orcidlink{0000-0002-1586-5790}} 
  \author{H.~Atmacan\,\orcidlink{0000-0003-2435-501X}} 
% ...
\collaboration{The Belle II Collaboration}

\begin{abstract}
The ratio of branching fractions $R(D^{*}) = \mathcal{B}(\overline{B} \rightarrow D^{*} \tau^{-} \overline{\nu}_{\tau})$/$\mathcal{B} (\overline{B} \rightarrow D^{*} \ell^{-} \overline{\nu}_{\ell})$, where $\ell$ is an electron or muon, is measured using a Belle~II data sample with an integrated luminosity of $189~\mathrm{fb}^{-1}$ at the SuperKEKB asymmetric-energy $e^{+} e^{-}$ collider. Data is collected at the $\Upsilon(\mathrm{4S})$ resonance, and one $B$ meson in the  $\Upsilon(\mathrm{4S})\rightarrow B\overline{B}$ decay is fully reconstructed in hadronic decay modes. The accompanying signal $B$ meson is reconstructed as $\overline{B}\rightarrow D^{*} \tau^{-}\overline{\nu}_{\tau}$ using leptonic $\tau$ decays. The normalization decay, $\overline{B}\rightarrow D^{*} \ell^{-} \overline{\nu}_{\ell}$, where $\ell$ is an electron or muon, produces the same observable final state particles. The ratio of branching fractions is extracted in a simultaneous fit to two signal-discriminating variables in both channels and yields $R(D^{*}) = 0.262~_{-0.039}^{+0.041}(\mathrm{stat})~_{-0.032}^{+0.035}(\mathrm{syst})$. This result is consistent with the current world average and with standard model predictions. 
\end{abstract} 

\maketitle

\section{Introduction}
\label{sec:intro}

In the Standard Model (SM) of particle physics, semileptonic $B$ decays proceed via $b \rightarrow c$ or $b \rightarrow u$ transitions and are mediated by a $W$ boson to produce a charged lepton and its corresponding neutrino. The decay rate of $\overline{B} \rightarrow D^{(*)}\ell^{-} \overline{\nu}_{\ell}$~\footnote{Charge-conjugate modes are implied throughout the paper.} involves the magntiude of the Cabibbo-Kobayashi-Maskawa matrix element $V_{cb}$. Hadronic effects that describe the nonperturbative physics of the $B \rightarrow D^{*}$ transition are also included in the decay rate and are described by hadronic matrix elements. The latter are parametrized, in the context of the heavy quark effective theory (HQET), in terms of form factors.

The $W$ boson couples equally to the three lepton generations~\cite{HFLAV, LEP_LU, ATLAS_LU, CMS_LU}, a symmetry known as lepton flavor universality (LFU). The LFU symmetry is a fundamental postulate of the SM and can be tested by measuring
\begin{equation}
R(D^{(*)})=\frac{\mathcal{B}(\overline{B} \rightarrow D^{(*)}\tau^{-} \overline{\nu}_{\tau})}
 {\mathcal{B}(\overline{B} \rightarrow D^{(*)} \ell^{-} \overline{\nu}_{\ell})},
\end{equation}
where the denominator is referred to as the normalization mode with $\ell=e$ or $\mu$. Semileptonic $B$ decays involving a $\tau$ lepton are sensitive to physics beyond the SM (BSM)~\cite{NP1, NP2, NP3, London_2022} and their decays are less constrained by data than semileptonic decays to electrons and muons. While the coupling to all lepton flavors is the same in the SM, the large value of the $\tau$ mass results in a reduced phase space factor, and hence $R(D)$ and $R(D^{*})$ are expected to be $0.298 \pm 0.004$ and $0.254 \pm 0.005$, respectively~\cite{HFLAV}.

In the $R(D^{(*)})$ ratios, $|V_{cb}|$ cancels, as do many of the theoretical and experimental uncertainties, such as the uncertainty on the Belle~II data set size. The cancellations make these measurements stringent LFU tests, which are often combined.

The $R(D^{(*)})$ ratios have been measured by the BaBar~\cite{BabarDtaunu1, BabarDtaunu2}, Belle~\cite{BelleDtaunu1, BelleDtaunu3, BelleDtaunu4}, and LHCb~\cite{LHCbDtaunu, LHCbDtaunu2, LHCb2021Paper, LHCb2023, LHCb2023_2} collaborations. The world averages of these measurements, $R(D) = 0.356 \pm 0.029$ and $R(D^{*}) = 0.284 \pm 0.013$, exceed the SM expectation by $2.0\sigma$ and $2.2\sigma$, respectively. The deviation in ($R(D)$, $R(D^{*})$) reaches $3.2\sigma$~\cite{HFLAV}. A recent measurement of the inclusive ratio, $R(X) = \mathcal{B}(\overline{B} \rightarrow X\tau^{-}\overline{\nu}_{\tau}) / \mathcal{B}(\overline{B} \rightarrow X\ell^{-}\overline{\nu}_{\ell})$, where $X$ denotes a hadronic system inclusively, is consistent with both the SM expectation and the $R(D^{(*)})$ averages~\cite{Belle2:2023RX}

Here we report the first measurement of $R(D^{*})$ at the Belle~II experiment, using a $189~\mathrm{fb}^{-1}$ sample of electron-positron collisions, corresponding to $N_{B\overline{B}} = (198.0 \pm 3.0) \times 10^{6}$ $B\overline{B}$ pairs, collected at the $\Upsilon(\mathrm{4S})$ resonance during the 2019--2021 run period. One $B$ meson, hereafter referred to as $B_{\mathrm{tag}}$, is fully reconstructed via hadronic decay modes and the remaining particles in the event are used to reconstruct the pair-produced signal, $\overline{B} \rightarrow D^{*}\tau^{-}\overline{\nu}_{\tau}$, and normalization mode decays, $\overline{B} \rightarrow D^{*}\ell^{-}\overline{\nu}_{\ell}$. Only leptonic $\tau$ decays are considered: $\tau^{-} \rightarrow e^{-}\overline{\nu}_e\nu_{\tau}$ and $\tau^{-} \rightarrow \mu^{-}\overline{\nu}_{\mu}\nu_{\tau}$. We extract the value of $R(D^{*})$ using a two-dimensional fit to two variables: the missing mass squared, $M_{\mathrm{miss}}^{2}$, and the residual calorimeter energy, $E_{\mathrm{ECL}}$. The definition of $M_{\mathrm{miss}}^{2}$ is given by
\begin{equation}
    M_{\mathrm{miss}}^{2}
     =\left(E_{\mathrm{beam}}^{*}-E_{D^{*}}^{*}-E_{\ell}^{*}\right)^2-(-\vec{p}_{B_{\mathrm{tag}}}^{~*}-\vec{p}_{D^{*}}^{~*}-\vec{p}_{\ell}^{~*})^{2}.
\end{equation}
Here $E_{\mathrm{beam}}^{*} = \sqrt{s} / 2$ represents the center-of-mass (c.m.) beam energy whereas $E_{B_{\mathrm{tag}}}^{*}$ ($\vec{p}_{B_{\mathrm{tag}}}^{~*}$), $E_{D^{*}}$ ($\vec{p}_{D^{*}}^{~*}$), and $E_{\ell}$ ($\vec{p}_{\ell}^{~*}$) are the energies (momentum three-vectors) of the $B_{\mathrm{tag}}$, $D^{*}$, and $\ell$, respectively, in the c.m.\ frame. The $E_{\mathrm{ECL}}$ quantity is defined as the linear sum of the energies detected in the calorimeter not associated with the reconstructed $B\overline{B}$ pair.

\section {Belle~II Experiment}
\label{sec:belle2}

The Belle~II detector~\cite{belle2tdr} is a general-purpose detector located at the asymmetric-energy collider accelerator complex SuperKEKB \cite{Akai:2018mbz}, where 7-GeV electrons collide with 4-GeV positrons at the c.m. energy of $10.58~\mathrm{GeV}$. This energy corresponds to the $\Upsilon(\mathrm{4S})$ resonance, which almost always decays to a $B\overline{B}$ pair. Belle~II uses cylindrical coordinates in which the $z$ axis is aligned along the solenoid axis and points approximately in the direction of the electron beam. The detector itself consists of seven main subdetector components and a superconducting solenoid that provides a magnetic field of $1.5~\mathrm{T}$.

Trajectories of charged particles (tracks) passing through a given detector region, along with their corresponding momenta and electric charge, are determined by the Belle II tracking system. It consists of three components: the pixel detector (PXD), the silicon vertex detector (SVD), and the central drift chamber (CDC). The PXD is closest to the interaction point (IP) and consists of two layers of high-granularity pixel sensors. For the data used in this measurement, only the innermost PXD layer, and one sixth of the outermost layer are installed. It is surrounded by the SVD, which is composed of four layers of double-sided silicon strip detectors. The PXD and SVD provide precise measurements of decay vertices. The CDC is a wire chamber filled with a $50\%$-$50\%$ mixture of helium and ethane. It surrounds the SVD and provides measurements of the momenta and ionization energy loss of charged particles. Outside the CDC, a quartz-based time-of-propagation counter and a proximity-focusing aerogel ring-imaging Cherenkov detector are located in the barrel and the forward endcap regions, respectively. These two detectors identify hadrons by reconstructing the timing and spatial structure of ring images of Cherenkov light cones.

Further out is the electromagnetic calorimeter (ECL) which is composed of CsI(Tl) scintillator crystals to measure the energy deposits, referred to as clusters, with their timing information. The ECL information is used mainly to reconstruct photons and distinguish electrons from other charged particles. The subdetectors described above are immersed in the magnetic field provided by the superconducting solenoid. Outside of the solenoid, a subdetector dedicated to identifying $K_{L}^{0}$ mesons and muons is installed. It consists of an alternating structure of $4.7~\mathrm{cm}$ thick iron plates and active detector elements. These detector elements consist of layers of either scintillator plates in the inner part of the barrel region and the endcaps or resistive-plate chambers in the outer part of the detector's barrel region~\cite{Aushev:2014spa}.

\section{Data and Simulation}
\label{sec:dataMC}

Monte Carlo (MC) simulation samples are used to develop the signal selection criteria, examine the leading background processes, and determine the fit model. The decay chains are simulated using the EvtGen package~\cite{Lange:2001uf} and the detector response is modeled with the Geant4 framework~\cite{geant4}. These samples consist of either $e^{+}e^{-} \rightarrow B\overline{B}$ events where each $B$ meson decays inclusively or continuum events. The decay rates of $B$ meson decays for which no measurements exist are modeled by PYTHIA~\cite{Sjostrand:2007gs}. Continuum events, defined as $e^{+}e^{-} \rightarrow q\overline{q}$ decays where $q$ is a $u$, $d$, $c$, or $s$ quark, are simulated with the KKMC package~\cite{kkmc} using PYTHIA~\cite{Sjostrand:2007gs} for hadronization. For all simulated events, electromagnetic final-state radiation is simulated using the PHOTOS package~\cite{PHOTOS, PHOTOS_new}. The corresponding luminosity of the $B\overline{B}$ and continuum samples is $0.9~\mathrm{ab}^{-1}$ and $1.0~\mathrm{ab}^{-1}$, respectively. A sample of 800 million simulated signal events is also generated, where one $B$ meson decays exclusively to $D^{(*)} \tau \nu_{\tau}$ and the other $B$ meson decays inclusively. All data and simulated events are analyzed using the Belle~II analysis software framework~\cite{basf2, basf2Zenodo}.

We correct the branching fractions of the $D$ meson decays used in the simulation to match the known values~\cite{pdg:2022}. Additionally, the branching fractions of the hadronic $B$ decays $\overline{B} \rightarrow D^{*}\overline{D}{}_{s}^{(*)}$, $\overline{B} \rightarrow D^{*}\overline{D}{}^{(*)}K$, and $\overline{B} \rightarrow D^{*}\mathrm{n}\pi(\pi^{0})$ are also corrected, where $\mathrm{n}$ indicates charged-pion multiplicity.

The heavier 1P charmed mesons, collectively known as $D^{**}$, are a leading background in this measurement and their description in the simulation is thus a critical component. The $D^{**}$ states predominantly decay to $D^{(*)}\mathrm{n}\pi$ states with multiplicity $\mathrm{n} > 0$. According to HQET, there are two narrow resonant $D^{**}$ states, $D_{1}$(2420) and $D_{2}^{*}$(2460), with a decay width of approximately $20~\mathrm{MeV}$, and two broad resonant $D^{**}$ states, $D_{0}^{*}$(2400) and $D_{1}'$(2430), with widths of $\mathcal{O}(100)~\mathrm{MeV}$. 

In the simulation, an isospin factor of $2/3$ is used to compute the overall branching fraction of $D^{**}$ decays to two-body final states, $D^{(*)}\pi$. The simulated average branching fraction of $D_{1}'$ excludes the result reported in Ref.~\cite{BelleDstst} that disagrees with the measurements of Refs.~\cite{BaBarDstst, DELPHIDstst}. The $\overline{B} \rightarrow D_{2}^{*}\ell^{-}\overline{\nu}_{\ell}$ branching fractions are computed using the $\mathcal{B}(D_{2}^{*} \rightarrow D\pi^{-}) / \mathcal{B}(D_{2}^{*} \rightarrow D^{*}\pi^{-})$ observed average~\cite{HFLAV}. As $D_{1}$ decays to both three-body final states, $D\pi\pi$, and two-body final states, ${D^{*}}\pi$, we estimate the full $\overline{B} \rightarrow D_{1}\ell^{-}\overline{\nu}$ branching fractions using measurements of the partial branching fractions with two-body $D_{1}$ final states~\cite{HFLAV}. This estimate depends on the branching fraction ratio $\mathcal{B}(D_{1}^{0} \rightarrow D^{+}\pi^{-})/\mathcal{B}(D_{1}^{0} \rightarrow D^{0}\pi^{+}\pi^{-})$~\cite{LHCb:D3pi} and the isospin factor of $\mathcal{B}(D^{**} \rightarrow D^{(*)}\pi^{+}\pi^{-}) / \mathcal{B}(D^{**} \rightarrow D^{(*)}\pi\pi) = 1/2 \pm 1/6$, where the uncertainty arises from the contribution of $\rho$ and $f_{0}$ resonances.

The nonresonant components $\overline{B} \rightarrow D^{(*)}\pi\pi\ell^{-}\overline{\nu}_{\ell}$ are simulated assuming they come from the broad $D_{0}^{*}$ and $D_{1}^{'}$ resonances in equal fractions. The branching fractions are derived using $\mathcal{B}(\overline{B} \rightarrow D^{(*)}\pi\pi\ell^{-}\overline{\nu}_{\ell})/\mathcal{B}(\overline{B} \rightarrow D^{(*)}\ell^{-}\overline{\nu}_{\ell})$~\cite{BaBarDstpipi}. For nonresonant $\overline{B} \rightarrow D^{(*)}\pi\ell^{-}\overline{\nu}_{\ell}$ decays, the branching fractions are consistent with zero once the resonant decay contributions are subtracted from the inclusive branching fractions for $\overline{B} \rightarrow D^{(*)}\pi\ell^{-}\overline{\nu}_{\ell}$. Consequently, these nonresonant branching fractions are set to zero. However, these nonresonant contributions are taken into account in the evaluation of the systematic uncertainties related to the composition of the $\overline{B} \rightarrow D^{**}\ell^{-}\overline{\nu}_{\ell}$ component. 
Finally, there is a gap between the branching fraction of the inclusive $\overline{B} \rightarrow X_{c}\ell^{-}\overline{\nu}_{\ell}$ decay and the sum of exclusive semileptonic $B$ decays to a charm meson. The decays $\overline{B} \rightarrow D_{0}^{*}(\rightarrow D\eta)\ell^{-}\overline{\nu}_{\ell}$ and $\overline{B} \rightarrow D_{1}'(\rightarrow D^{*}\eta)\ell^{-}\overline{\nu}_{\ell}$ are assigned to fill this branching ratio gap and an uncertainty of $100\%$ is assumed in the systematic uncertainty evaluation. We collectively refer to the gap component as $\overline{B} \rightarrow D_{\mathrm{gap}}^{**}\ell^{-}\overline{\nu}_{\ell}$.
As $\overline{B} \rightarrow D^{**}\tau^{-}\overline{\nu}_{\tau}$ decays have not been observed, we calculate their branching fractions using those of $\overline{B} \rightarrow D_{(\mathrm{gap})}^{**}\ell^{-}\overline{\nu}_{\ell}$ and assuming $R(D^{**}) = \mathcal{B}(\overline{B}\rightarrow D^{**}\tau^{-}\overline{\nu}_{\tau}) / \mathcal{B}(\overline{B}\rightarrow D^{**}\ell^{-}\overline{\nu}_{\ell}) = 0.085 \pm 0.012$~\cite{BLR2}. An uncertainty of $100\%$ is also assigned to the branching fractions of $\overline{B} \rightarrow D^{**}\tau^{-}\overline{\nu}_{\tau}$ decays.

\begin{table*}
    \centering
    \caption{
        Simulated branching fractions of $\overline{B} \rightarrow D^{**}\ell^{-}\overline{\nu}$ decays used for modeling the leading background. The branching fractions used for the evaluation of the systematic uncertainty due to nonresonant $\overline{B} \rightarrow D^{(*)}\pi\ell^{-}\overline{\nu}_{\ell}$ are shown in parentheses.
    }
    \begin{tabular}{lcc}
        \toprule
        \toprule
        \multicolumn{1}{c}{\multirow{2}{*}{Decay}} & \multicolumn{2}{c}{\hspace{1.0cm} Branching fraction~$[10^{-3}]$ \hspace{1.0cm} } \\
        \cmidrule{2-3}
        & \multicolumn{1}{c}{$\mathcal{B}(B^{0})$} & \multicolumn{1}{c}{$\mathcal{B}(B^{+})$} \\
        \midrule
        $\overline{B} \rightarrow D_{1}\ell^{-}\overline{\nu}_{\ell}$ \hspace{1.0cm} & $6.16 \pm 1.01$ & $6.63 \pm 1.09$ \\
        $\overline{B} \rightarrow D_{0}^{*}\ell^{-}\overline{\nu}_{\ell}$ & $3.90 \pm 0.70$ & $4.20 \pm 0.75$ \\
        $\overline{B} \rightarrow D_{1}'\ell^{-}\overline{\nu}_{\ell}$ & $3.90 \pm 0.84$ & $4.20 \pm 0.90$ \\
        $\overline{B} \rightarrow D_{2}^{*}\ell^{-}\overline{\nu}_{\ell}$ & $2.73 \pm 0.30$ & $2.93 \pm 0.32$ \\
        \midrule
        $\overline{B} \rightarrow D_{s}K\ell^{-}\overline{\nu}_{\ell}$ & --- & $0.30 \pm 0.14$ \\
        $\overline{B} \rightarrow D_{s}^{*}K\ell^{-}\overline{\nu}_{\ell}$ & --- & $0.29 \pm 0.19$ \\
        \midrule
        $\overline{B} \rightarrow D\pi\ell^{-}\overline{\nu}_{\ell}$ & $0$~~~($0.3 \pm 0.9$) & $0$~~~($0.3 \pm 0.9$) \\
        $\overline{B}\rightarrow D^{*}\pi\ell^{-}\overline{\nu}_{\ell}$ & $0$~($-1.1 \pm 1.1$) & $0$~($-1.1 \pm 1.1$) \\
        $\overline{B} \rightarrow D\pi\pi\ell^{-}\overline{\nu}_{\ell}$ & $0.58 \pm 0.82$ & $0.62 \pm 0.89$ \\
        $\overline{B} \rightarrow D^{*}\pi\pi\ell^{-}\overline{\nu}_{\ell}$ & $2.01 \pm 0.95$ & $2.16 \pm 1.02$ \\
        $\overline{B} \rightarrow D\eta\ell^{-}\overline{\nu}_{\ell}$ & $4.09 \pm 4.09$ & $3.77 \pm 3.77$ \\
        $\overline{B} \rightarrow D^{*}\eta\ell^{-}\overline{\nu}_{\ell}$ & $4.09 \pm 4.09$ & $3.77 \pm 3.77$ \\
        \midrule
        $\overline{B} \rightarrow D_{1}\tau^{-}\overline{\nu}_{\tau}$ & $0.52 \pm 0.52$ & $0.56 \pm 0.56$ \\
        $\overline{B} \rightarrow D_{0}^{*}\tau^{-}\overline{\nu}_{\tau}$ & $0.33 \pm 0.33$ & $0.36 \pm 0.36$ \\
        $\overline{B} \rightarrow D_{1}'\tau^{-}\overline{\nu}_{\tau}$ & $0.33 \pm 0.33$ & $0.36 \pm 0.36$ \\
        $\overline{B} \rightarrow D_{2}^{*}\tau^{-}\overline{\nu}_{\tau}$ & $0.23 \pm 0.23$ & $0.25 \pm 0.25$ \\
        \midrule
        $\overline{B} \rightarrow D\pi\pi\tau^{-}\overline{\nu}_{\tau}$ & $0.05 \pm 0.05$ & $0.05 \pm 0.05$ \\
        $\overline{B} \rightarrow D^{*}\pi\pi\tau^{-}\overline{\nu}_{\tau}$ & $0.17 \pm 0.17$ & $0.18 \pm 0.18$ \\
        $\overline{B} \rightarrow D\eta\tau^{-}\overline{\nu}_{\tau}$ & $0.35 \pm 0.35$ & $0.32 \pm 0.32$ \\
        $\overline{B} \rightarrow D^{*}\eta\tau^{-}\overline{\nu}_{\tau}$ & $0.35 \pm 0.35$ & $0.32 \pm 0.32$ \\
        \bottomrule
        \bottomrule
    \end{tabular}
    \label{tab:DststlnuBRs}
\end{table*}

The branching fractions into resonant $D^{**}\ell^{-}\overline{\nu}_{\ell}$ and $D^{**}\tau^{-}\overline{\nu}_{\tau}$, and nonresonant $D_{s}^{(*)}K\ell^{-}\overline{\nu}_{\ell}$, $D^{(*)}\pi\ell^{-}\overline{\nu}_{\ell}$, and $D^{(*)}\pi\pi\ell^{-}\overline{\nu}_{\ell}$, $D^{(*)}\eta\ell^{-}\overline{\nu}_{\ell}$, $D^{(*)}\pi\pi\tau^{-}\overline{\nu}_{\tau}$, and $D^{(*)}\eta\tau^{-}\overline{\nu}_{\tau}$ final states implemented in the simulation samples are listed in Table~\ref{tab:DststlnuBRs}. We assume that isospin symmetry holds in these decays and correct the branching fractions in the simulation according to the isospin-averaged branching fractions. The $\overline{B} \rightarrow D^{**}\ell^{-}\overline{\nu}_{\ell}$ events are thus weighted by the corrected branching fractions for both neutral and charged $B$ mesons.

The simulation uses a parametrization of the hadronic form factors for semileptonic $B$ decays based on HQET. Both simulated signal and normalization samples are based on the form factor parametrization of Ref.~\cite{BLPRXP}. We follow Refs.~\cite{BLR1, BLR2} to model $\overline{B} \rightarrow D^{**}\ell^{-}\overline{\nu}_{\ell}$ and $\overline{B} \rightarrow D^{**}\tau^{-}\overline{\nu}_{\tau}$ decays. These form factors are updated to their most recent values by determining event-by-event weights and applying them to final distributions using HAMMER~\cite{Hammer}.

\section{Reconstruction and Event Selection}
\label{sec:rec}

\subsection{Selection of $\Upsilon(4S)$ events}

Events are selected only based on online trigger criteria that count the number of tracks or ECL clusters, or the summed energy of all clusters. These trigger selections have nearly $100\%$ efficiency. In the offline analysis, an optimized selection is applied to the tracks and clusters in a given event to select $\overline{B} \rightarrow D^{*}\tau^{-}\overline{\nu}_{\tau}$ and $\overline{B} \rightarrow D^{*}\ell^{-}\overline{\nu}_{\ell}$ decays. All events are required to have at least five tracks and at least three clusters, where the latter includes clusters associated with tracks.
The impact parameters of tracks with respect to the IP must be less than $2~\mathrm{cm}$ along the $z$ axis, and $0.5~\mathrm{cm}$ transverse to the $z$ axis. The minimum accepted transverse momentum, $p_{\mathrm{T}}$, is $100~\mathrm{MeV}/c$ for all charged particles except the low-momentum pion daughters of the ${D^{*}}^{+}$ mesons, for which the requirement is $50~\mathrm{MeV}/c$. To exclude events from two-photon processes, we require the measured visible energy, $E_{\mathrm{vis}}$, defined as the sum of all the measured energies of the charged particles and neutral clusters, which are clusters not associated with tracks, in the event, to be greater than $4~\mathrm{GeV}$.
Continuum events are suppressed by requiring $R_{2} < 0.4$, where $R_{2}$ is the ratio of the second to zeroth Fox-Wolfram moments~\cite{Fox:1978vu}, which is a measure of the sphericity of the spatial distribution of final-state particles. In the calculation of $E_{\mathrm{vis}}$ and $R_{2}$, only neutral clusters with an energy of at least $100~\mathrm{MeV}$ and an associated polar angle within the CDC acceptance region ($17^{\circ} < \theta < 150^{\circ}$ in the laboratory frame) are included. 

\subsection{Reconstruction of tag side $B$ meson}

Simulation samples and collision data are initially passed through the full event interpretation (FEI)~\cite{fei}, a hierarchical multivariate algorithm that fully reconstructs one of the $B$ mesons in a hadronic decay mode. The output of the FEI algorithm is a list of $B_{\mathrm{tag}}$ candidates with a probability ranging between zero and one, with zero (one) corresponding to a low (high) probability that the $B_{\mathrm{tag}}$ candidate is properly reconstructed. For this measurement, $B_{\mathrm{tag}}$ candidates are required to have a FEI probability greater than $0.01$. Furthermore, the selected $B_{\mathrm{tag}}$ candidate must satisfy $M_{\mathrm{bc}} > 5.27~\mathrm{GeV}/c^{2}$ and $-0.15 < \Delta E <0.1~\mathrm{GeV}$. Here, $M_{\mathrm{bc}}$ is defined as 
\begin{equation}
    M_{\mathrm{bc}}=\sqrt{\left(E_{\mathrm{beam}}^{*}\right)^{2} - |\vec{p}_{B_{\mathrm{tag}}}^{~*}c|^2} / c^{2}.
\end{equation}
The variable $\Delta E = E^{*}_{B_{\mathrm{tag}}} - E_{\mathrm{beam}}^{*}$ is the difference between the observed c.m.\ energy of the $B_{\mathrm{tag}}$ candidate and its expected value $E_{\mathrm{beam}}^{*}$. The resulting fraction of $\Upsilon(\mathrm{4S})$ events with a correctly reconstructed $B_{\mathrm{tag}}$ candidate is approximately $0.23\%$ for $B^{0}$ and $0.30\%$ for $B^{+}$ with a purity of $29\%$~\cite{FEICalib}. 

\subsection{Reconstruction of the signal $B$ meson}
\label{rec_sigB}

The signal $B$, referred to as $B_{\mathrm{sig}}$, is reconstructed with the following combinations of a $D^{*}$ meson and a lepton candidate: $(D^{*}{}^{+}, e^{-})$, $(D^{*}{}^{+}, \mu^{-})$, $(D^{*}{}^{0}, e^{-})$, and $(D^{*}{}^{0}, \mu^{-})$. Candidate $D^{*}$ mesons are reconstructed in their $D^{0}\pi^{+}$, $D^{+}\pi^{0}$, and $D^{0}\pi^{0}$ decays. The $D^{+}$ candidates are reconstructed in the decay modes $K^{-} \pi^{+} \pi^{+}$, $K_{S}^{0} \pi^{+}$, and $K^{-} K^{+} \pi^{+}$. The $D^{0}$ candidates are reconstructed in the decay modes $K^{-} \pi^{+} \pi^{0}$, $K^{-} \pi^{+}  \pi^{-}\pi^{+}$, $K_{S}^{0} \pi^{+} \pi^{-} \pi^{0}$,  $K^{-} \pi^{+}$,  $K_{S}^{0} \pi^{+} \pi^{-}$, $K_{S}^{0} \pi^{0}$, $K^{-} K^{+}$, and $\pi^{-} \pi^{+}$.
Kaon and pion candidates are required to have more than 20 measurement points (hits) in the CDC. Charged kaon candidates are required to satisfy the $\mathcal{P}_{K}$ = $\mathcal{L}_{K}/(\mathcal{L}_{K} + \mathcal{L}_{\pi}) > 0.1$ particle identification (PID) criterion, where $\mathcal{L}_{K}$ and $\mathcal{L}_{\pi}$ indicate the identification likelihoods for a kaon and pion, respectively. The identification likelihood for a pion or kaon hypothesis combines PID information from all subdetectors except the PXD. Charged pion candidates, except for the low-momentum pion daughter of the ${D^{*}}^{+}$, are required to satisfy $\mathcal{P}_{\pi} = \mathcal{L}_{\pi}/(\mathcal{L}_{K} + \mathcal{L}_{\pi}) > 0.1$. These PID selection criteria discriminate kaons and pions with efficiencies of $94.1\%$ and $97.5\%$ at misidentification rates of $14.0\%$ and $7.3\%$, respectively.

Photon candidates must lie within the angular acceptance of the CDC, and satisfy polar-angle-dependent energy requirements: $E > 80~\mathrm{MeV}$, $30~\mathrm{MeV}$, and $60~\mathrm{MeV}$ in the forward endcap, barrel, and backward endcap regions of the ECL, respectively. Candidate $\pi^0$'s are reconstructed via $\pi^{0} \rightarrow \gamma\gamma$ decays, where additional requirements are applied on the photons to further reduce background from misreconstructed candidates. Requirements on the distance of each photon to the nearest track extrapolated to the ECL, denoted as $\Delta_{\mathrm{TC}}$, and on an electromagnetic shower shape-based classifier variable~\cite{Longo:ZernikeMVA}, denoted as $Z_{0}$ and determined using 11 Zernike moments~\cite{Zernike, Vasudevan:Zernike} in the ECL, are applied. The latter quantity is used to distinguish between clusters generated by real photons and those that result from $K_{L}^{0}$ mesons or hadronic showers. The photon candidates are selected by applying an optimized requirement based on the relation $(\Delta_{\mathrm{TC}}/X)^{2} + (Z_{0}/Y)^{2} > 1$ for each $D^{(*)}$ decay mode and for each of the forward endcap, barrel, and backward endcap regions. The values of $X$ and $Y$ are chosen to maximize the figure of merit (FOM), defined as $N_{\mathrm{sig}}/\sqrt{N_{\mathrm{sig}} + N_{\mathrm{bkg}}}$, where $N_{\mathrm{sig}}$ $(N_{\mathrm{bkg}})$ is the number of signal (background) events obtained from simulation in the $M_{\mathrm{miss}}^{2} > 0.5~\mathrm{GeV}^{2}/c^{4}$ region. The mass of the reconstructed $\pi^{0}$ candidates must lie within the intervals [0.1224, 0.1430]~$\mathrm{GeV}/c^{2}$ and [0.1183, 0.1470]~$\mathrm{GeV}/c^{2}$ for the $D^{0}$ and $D^{*}$ daughters, respectively. Fractions of $94\%$ and $88\%$ of correctly reconstructed $\pi^{0}$ candidates pass the mass selections, respectively.

For the low-momentum $\pi^{0}$ daughter of the $D^{*}$, an additional requirement on the energy asymmetry, $A_{E} = (E_{\gamma_{\mathrm{high}}} - E_{\gamma_{\mathrm{low}}}) / (E_{\gamma_{\mathrm{high}}} + E_{\gamma_{\mathrm{low}}})$, where $E_{\gamma_{\mathrm{high}}}$ ($E_{\gamma_{\mathrm{low}}}$) are the larger (smaller) of the photon energies, is applied: $A_{E} < 0.65$. This selection reduces background $\pi^{0}$ candidates that arise from two-photon combinations involving a background photon. In addition, a less restrictive cluster-energy requirement, $E > 25~\mathrm{MeV}$ in the forward endcap and barrel ECL regions and $E > 40~\mathrm{MeV}$ in the backward endcap region, is applied for the low-momentum $\pi^{0}$ in ${D^{*}}^{0}$ decays.

Candidate $K_{S}^{0}$ mesons are reconstructed in their $K_{S}^{0} \rightarrow \pi^{-}\pi^{+}$ decays. We employ a FastBDT classifier~\cite{Keck:BDT} to discriminate $K_{S}^{0}$ candidates from $\pi^{+}\pi^{-}$ combinatorial background and $\Lambda^{0}$ baryon decays to $p\pi^{-}$. The classifier returns two $K_{S}^{0}$ probabilities, referred to as $P_{K_{S}^{0}}$ and $P_{\Lambda^{0}\mbox{-}\mathrm{veto}}$, respectively, utilizing the kinematic properties of the $K_{S}^{0}$ and its daughter pions, the flight length of the $K_{S}^{0}$, and the number of hits in PXD and SVD as input variables. We require $P_{K_{S}^{0}} > 0.90$ and $P_{\Lambda^{0}\mbox{-}\mathrm{veto}} > 0.11$. In addition, the reconstructed $K_{S}^{0}$ invariant mass is required to be between $0.4768$ and $0.5146~\mathrm{GeV}/c^{2}$ and its flight length from the IP less than $5.0~\mathrm{cm}$. The selection criteria for $K_{S}^{0}$ candidates have an efficiency of approximately $90\%$.

After the reconstruction of $D^{*}$ candidates, mode-dependent selection criteria are placed on the $D$ mass and the difference between the reconstructed $D^{*}$ and $D$ masses, $\Delta M_{D^{*}}= M_{D^{*}} - M_{D}$, to maximize the FOM. The optimized mass windows vary up to $5.0\sigma$ in width depending on the decay mode, where $\sigma$ denotes the mass resolution in a specific decay mode. Typically, the ranges are $2\sigma$--$3\sigma$, except for $\Delta M_{D^{*}}$ in the ${D^{*}}^{+} \rightarrow D^{+}(\rightarrow K^{-}K^{+}\pi^{+})\pi^{0}$ mode in which the window is $\pm 1.0\sigma$.

Each $D^{*}$, charged or neutral, is then combined with a lepton candidate to form the $B_{\mathrm{sig}}$ candidate. Lepton candidates are required to have a PID criterion above 0.9 for an electron or muon. The PID selection has an efficiency of $84.6\%$ and $87.7\%$ for electrons and muons at misidentification rates of $1.0\%$ and $4.8\%$, respectively. A vertex fit~\cite{VertexFit} is applied to the $D^{*}\ell$ combination, constraining the masses of all daughter $K_{S}^{0}$ and $\pi^{0}$ mesons to their known values. Any $B_{\mathrm{sig}}$ candidate that fails this fit is discarded. For all successful $B_{\mathrm{sig}}$ candidates, a secondary vertex fit is applied in which the masses of all meson daughters in the decay chain are constrained to improve the resolution of $M_{\mathrm{miss}}^{2}$. The $B_{\mathrm{sig}}$ candidates that fail the second fit are discarded.

The candidate $B_{\mathrm{sig}}$ is then combined with the $B_{\mathrm{tag}}$, including ($B_{\mathrm{sig}}^{0}$,  $\overline{B}{}_{\mathrm{tag}}^{0}$), ($B_{\mathrm{sig}}^{0}$,  $B_{\mathrm{tag}}^{0}$), ($B_{\mathrm{sig}}^{+}$,  $B_{\mathrm{tag}}^{-}$) combinations and their charge conjugates, to form an $\Upsilon(\mathrm{4S})$ candidate. We define the signal region as $q^{2} > 4.0~(\mathrm{GeV}/c)^{2}$. Here, $q^{2} = (p_{\tau/\ell} + p_{\nu_{\tau/\ell}})^{2} = (p_{B_{\mathrm{sig}}}-p_{D^{*}})^{2}$, where $p$ is the four-momentum of each particle and the $B_{\mathrm{sig}}$ momentum in the c.m. frame is assumed to be $-\vec{p}_{B_{\mathrm{tag}}}^{~*}$.

After the reconstruction of $\Upsilon(\mathrm{4S})$ candidates, all remaining tracks and clusters are attributed to the rest-of-the-event (ROE). Any $\Upsilon(\mathrm{4S})$ candidate with one or more residual ROE tracks with at least one hit in the CDC and a distance of closest approach less than $10.0~\mathrm{cm}$ ($5.0~\mathrm{cm}$) along (transverse to) the $z$ axis, is discarded. Neutral pion candidates are reconstructed using ROE clusters and are referred to as $\pi_{\mathrm{ROE}}^{0}$. The following selection criteria are applied to daughter clusters of $\pi_{\mathrm{ROE}}^{0}$: $17^{\circ} < \theta < 150^{\circ}$, and $E > 120~\mathrm{MeV}$, $30~\mathrm{MeV}$, and $80~\mathrm{MeV}$ in the forward endcap, barrel, and backward endcap regions of the ECL, respectively. In addition, for each ROE cluster, the ratio of the deposited energy in the central ECL crystal to the sum of the energies in the surrounding $3\times3$ set of ECL crystals must exceed $0.4$. If any $\pi_{\mathrm{ROE}}^{0}$ candidate has an invariant mass within [0.121, 0.142]~$\mathrm{GeV}/c^{2}$, and an angle between the momenta and transverse momenta of the photon daughters less than 1.0 and 0.9, respectively, the event is discarded. After these selection criteria, all residual neutral clusters in the ROE are used to calculate $E_{\mathrm{ECL}}$. These neutral clusters in the ROE  must satisfy the same requirements as those used to reconstruct the $\pi^{0}$ candidates in $D$ decays on the signal side with $\Delta_{\mathrm{TC}} > 20~\mathrm{cm}$.

\subsection{Candidate selection}

The $\Upsilon(\mathrm{4S})$ reconstruction process can result in accepting more than one candidate per event. A single candidate is chosen in each event as follows. The $B_{\mathrm{tag}}$ with the highest FEI signal probability is taken and the other candidates are discarded. In simulation, the corresponding retention rate of the correctly reconstructed signal and normalization candidates is more than 99\% for every $D^{*}$ mode after this requirement. On the $B_{\mathrm{sig}}$ side, approximately $3.3\%$ of the reconstructed $B^{0}$ and $7.1\%$ of the reconstructed $B^{+}$ events still have two or three $D^{*}\ell^{-}$ candidates. If ${D^{*}}^{+}\ell^{-}$ candidates are reconstructed in both the $D^{0}\pi^{+}$ and $D^{+}\pi^{0}$ modes, only the candidates reconstructed as ${D^{*}}^{+}\rightarrow D^{0} \pi^{+}$ are retained. Furthermore, if multiple $B_{\mathrm{sig}}$ candidates have a ${D^{*}}^{+}$ candidate reconstructed via the ${D^{*}}^{+}\rightarrow D^{0} \pi^{+}$ mode, the one with the highest $\chi^{2}$ probability for the $B_{\mathrm{sig}}$ vertex fit is chosen. For both ${D^{*}}^{+}\rightarrow D^{+} \pi^{0}$ and ${D^{*}}^{0}\rightarrow D^{0} \pi^{0}$ modes, we select the $B_{\mathrm{sig}}$ candidate that yields the minimum $\chi^{2}(M_{\gamma\gamma})$ value. This $\chi^{2}(M_{\gamma\gamma})$ is calculated as $(M_{\gamma\gamma} - M_{\pi^{0}}^{\mathrm{PDG}})^{2}/ \sigma_{M_{\gamma\gamma}}^{2}$, where $M_{\gamma\gamma}$ denotes the invariant mass of the low-momentum $\pi^{0}$ candidate, $M_{\pi^{0}}^{\mathrm{PDG}}$ is the known $\pi^{0}$ mass~\cite{pdg:2022}, and $\sigma_{M_{\gamma\gamma}}$ represents the asymmetric standard deviation of $M_{\gamma\gamma}$ determined in simulation. For all modes, if more than one candidate is still present, the candidates reconstructed from the daughter $D$ decay with the highest branching fraction are chosen. After the candidate selection, 0.01\% and 1.3\% of reconstructed $B^{0}$'s and $B^{+}$'s, respectively, still have more than one candidate. At this point, candidates are chosen at random.

\section{Calibration and Model Validation}
\label{sec:carlib}

\subsection{Efficiency correction}

To correct discrepancies between data and simulation in the efficiency of the hadronic FEI algorithm used for $B_{\mathrm{tag}}$ reconstruction, correction factors have been introduced~\cite{FEICalib}. These are obtained from a fit to the lepton energy spectrum in $\overline{B} \rightarrow X_{c}\ell^{-}\overline{\nu}_{\ell}$ decays. The correction factors are obtained separately for $B^{0}$ and $B^{\pm}$ events, and are also separated by final state lepton flavor. The correction factors for the neutral (charged) modes are $0.710 \pm 0.023$ ($0.686 \pm 0.021$) and $0.673 \pm 0.025$ ($0.650 \pm 0.024$) for the electron and muon channels, respectively. 
Additional correction factors are applied to account for the following sources of mismodeling: the momentum scale of all charged particles in data;
the reconstruction efficiency for low-momentum charged particles;
the lepton identification (ID) efficiency and lepton ID misidentification rates;
the $K^{+}$ and $\pi^{+}$ ID efficiency; and
the $\pi^{0}$ efficiency. 
The $D^{(*)}$ meson mass resolutions are found to be different between data and simulation due to imperfections in the detector simulation. We therefore vary the width of the signal region, defined in both the $M_{D}$ and $\Delta M_{D^{*}}$ variables, with a correction factor for each decay mode to calibrate the yields of correctly reconstructed $D^{(*)}$ candidates in data. Mode-dependent correction factors are determined by a fit to the $M_{D}$ and $\Delta M_{D^{*}}$ distributions. These correction factors are typically within the $10\%$--$20\%$ range. The $D^{0} \rightarrow K_{S}^{0}\pi^{+}\pi^{-}$ mode requires the largest adjustment, at $43\%$.

\subsection{$\overline{B} \rightarrow D^{**}\ell^{-}\overline{\nu}_{\ell}$ composition}

\begin{figure}
    \centering\includegraphics[width=0.80\linewidth]{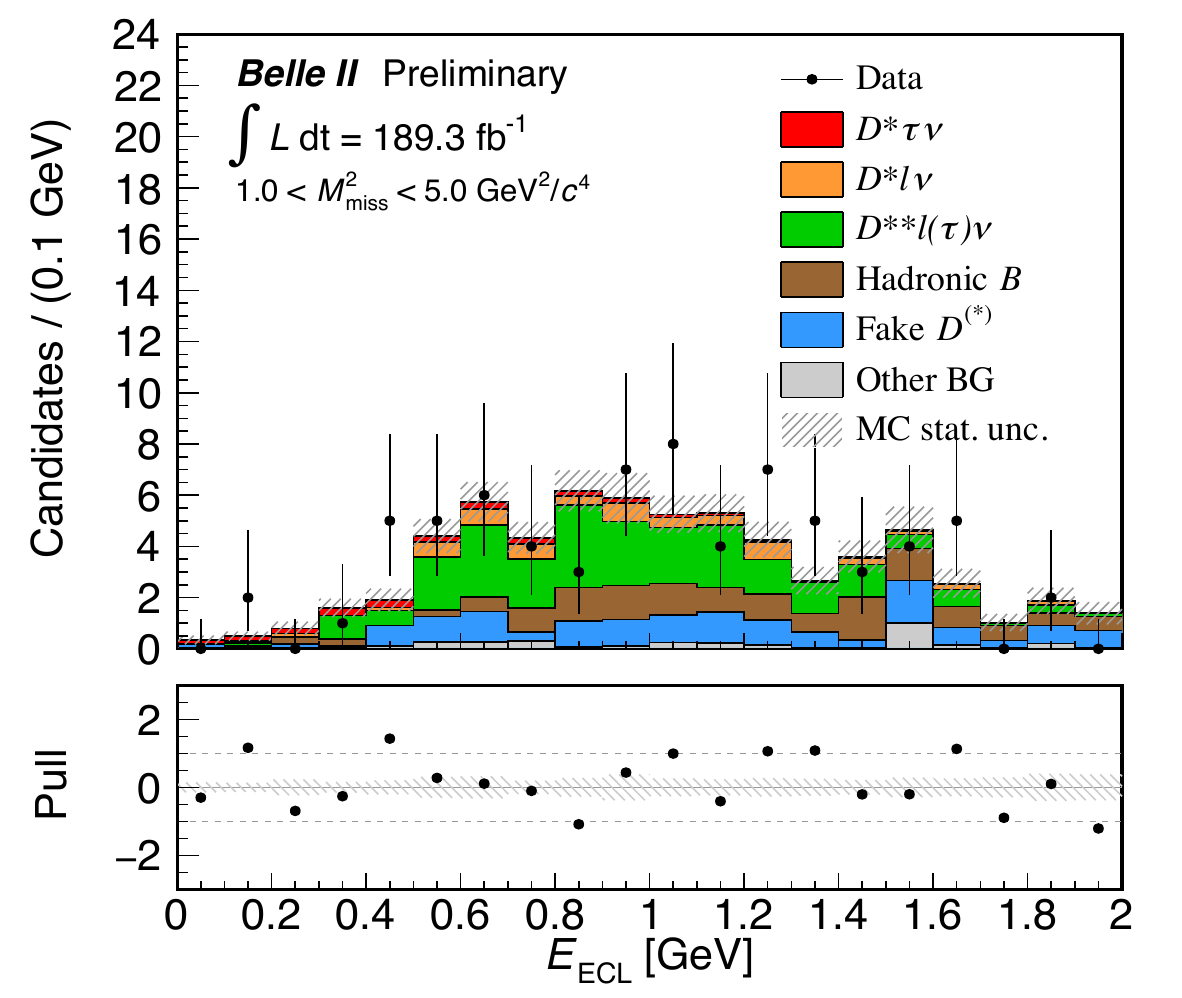}
    \caption{
        Comparison of the $E_{\mathrm{ECL}}$ distribution between data and simulation in the $\overline{B} \rightarrow D^{*}\pi^{0}\ell^{-}\overline{\nu}_{\ell}$ sideband region, where $1.0 < M_{\mathrm{miss}}^{2} < 5.0~\mathrm{GeV}^{2}/c^{4}$ is applied. The bottom panel shows the difference between data and simulation divided by the uncertainty in the data (pull).
    }
    \label{fig:EeclPi0ROESBXc100}
\end{figure}

A dedicated $\overline{B} \rightarrow D^{*}\pi^{0}\ell^{-}\overline{\nu}_{\ell}$ sideband, which is enhanced in $\overline{B}\rightarrow D^{**}\ell^{-}\overline{\nu}_{\ell}$ decays, is used to compare the $D^{**}$ yields in data and simulation. This includes the explicit reconstruction of additional neutral pions from $D^{**}$ decays, enabling a data-driven validation for the modeling of $\overline{B}\rightarrow D^{**}\ell^{-}\overline{\nu}_{\ell}$ decays in simulation. Therefore, we require at least one $\pi_{\mathrm{ROE}}^{0}$ candidate in the ROE.
The simulation is validated using the sideband data with $1.0 < M_{\mathrm{miss}}^{2} < 5.0~\mathrm{GeV}^{2}/c^{4}$, where the $\overline{B} \rightarrow D^{**}\ell^{-}\overline{\nu}_{\ell}$ events are dominant, as shown in Figure~\ref{fig:EeclPi0ROESBXc100}. The simulation reproduces the data well for all $D^{*}$ modes.

\subsection{Fake $D^{*}$ events}

After signal selection, the dominant source of background is from $B$ decays that contain misreconstructed $D^{*}$ meson candidates, denoted as fakes. They arise from two sources: a correctly reconstructed $D$ meson paired with a low-momentum pion that does not come from the same parent $D^{*}$ as the $D$ meson in question, and a misreconstructed $D$ meson paired with a low-momentum pion. 
Fake $D^{*}$ mesons are predominantly from $\overline{B} \rightarrow D^{*(*)}\ell^{-}\overline{\nu}$ decays, hadronic $B$ decays, ${B}^{0}\leftrightarrow B^{+}$ cross feed events where a  ${B}^{0}$ candidate is misreconstructed as $B^{+}$ and vice versa, and continuum events.

\begin{figure*}
    \centering
    \begin{tabular}{cc}
        \includegraphics[width=0.40\linewidth]{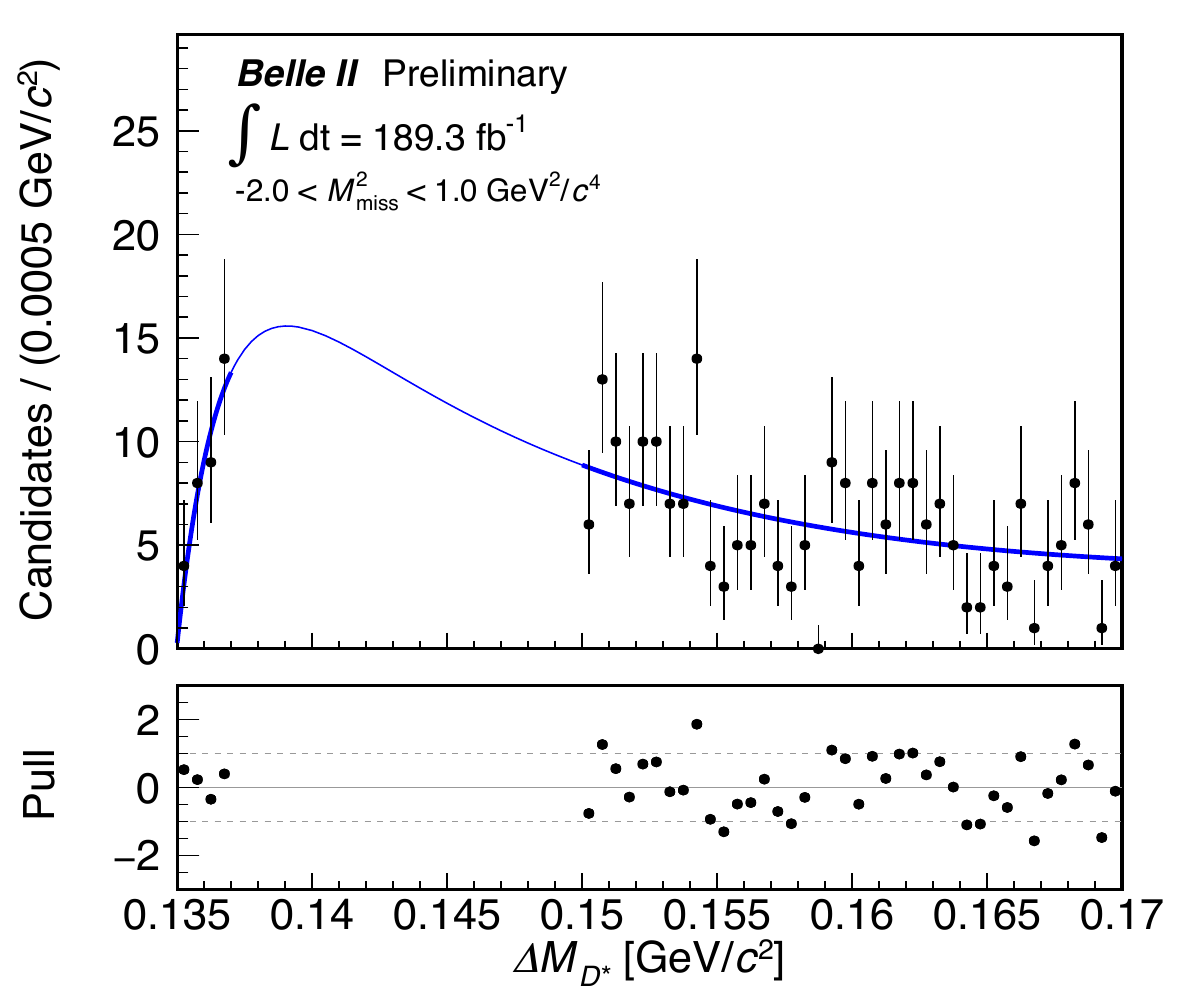} &
        \includegraphics[width=0.40\linewidth]{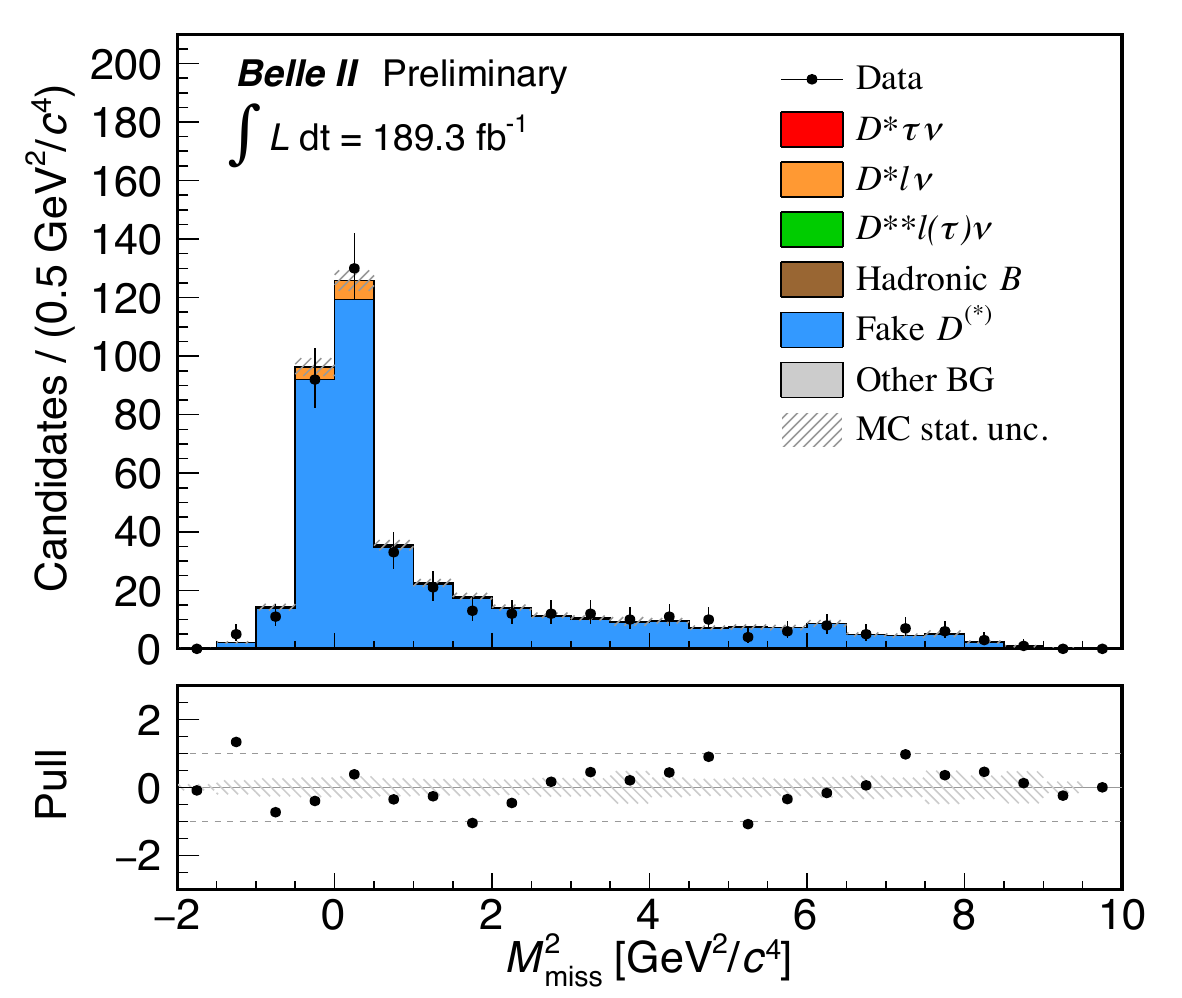}
    \end{tabular}
    \caption{
        Left: the distribution of $\Delta M_{D^{*}}$ (black points) in the sideband for ${D^{*}}^{0} \rightarrow D^{0}\pi^{0}$ at $M_{\mathrm{miss}}^{2} \in [-2.0, 1.0)~\mathrm{GeV}^{2}/c^{4}$ with fit results overlaid (blue line). Right: the distribution of $M_{\mathrm{miss}}^{2}$ in the $\Delta M_{D^{*}}$ region for data (black points) and simulation (histogram) after the calibration of the fake $D^{*}$ yields. The bottom panels show pull values from fit results or simulation.
    }
    \label{fig:fakeDstCalib}
\end{figure*}

The fake $D^{*}$ yields are calibrated from fits to the $\Delta M_{D^{*}}$ sideband region in data and simulation. The data-simulation ratio of yields of fake $D^{*}$ mesons are determined for each $D^{*}$ mode. The sideband region is defined as the $\Delta M_{D^{*}}$ regions [0.140, 0.141] and [0.155, 0.170]~$\mathrm{GeV}/c^{2}$ for ${D^{*}}^{+} \rightarrow D^{0}\pi^{+}$, and [0.135, 0.137] and [0.150, 0.170]~$\mathrm{GeV}/c^{2}$ for ${D^{*}}^{+} \rightarrow D^{+}\pi^{0}$ and ${D^{*}}^{0} \rightarrow D^{0}\pi^{0}$. The $\Delta M_{D^{*}}$ distributions are fitted using the threshold function
\begin{multline}
    f(\Delta M_{D^{*}}|M_{\pi}^{\mathrm{PDG}}, A, B, C) \\
    = \left(1 - \exp{\left(-\frac{\Delta M_{D^{*}} - M_{\pi}^{\mathrm{PDG}}}{A}\right)}\right) \times \left(\frac{\Delta M_{D^{*}}}{M_{\pi}^{\mathrm{PDG}}}\right)^{B} \\
    + C \left(\frac{\Delta M_{D^{*}}}{M_{\pi}^{\mathrm{PDG}}} - 1\right),
    \label{eq:fitThreFuncFakeD}
\end{multline}
where $M_{\pi}^{\mathrm{PDG}}$ is the known mass of the charged (neutral) pion for ${D^{*}}^{+} \rightarrow D^{0}\pi^{+}$ (${D^{*}}^{+} \rightarrow D^{+}\pi^{0}$ and ${D^{*}}^{0} \rightarrow D^{0}\pi^{0}$), and $A$, $B$, and $C$ are shape parameters determined from fake $D^{*}$ candidates in simulation. In the ${D^{*}}^{0} \rightarrow D^{0}\pi^{0}$ mode, to account for a small dependence of the data-simulation ratio on the $M_{\mathrm{miss}}^{2}$ distribution, the correction factors are determined and applied separately in three $M_{\mathrm{miss}}^{2}$ regions: $[-2.0, 1.0)$, $[1.0, 5.0)$, $[5.0, 10.0]~\mathrm{GeV}^{2}/c^{4}$.
The fit result and calibrated $M_{\mathrm{miss}}^{2}$ distribution are shown in Figure~\ref{fig:fakeDstCalib}.

\subsection{$E_{\mathrm{ECL}}$ shape}

\begin{figure}
    \centering
    \includegraphics[width=0.80\linewidth]{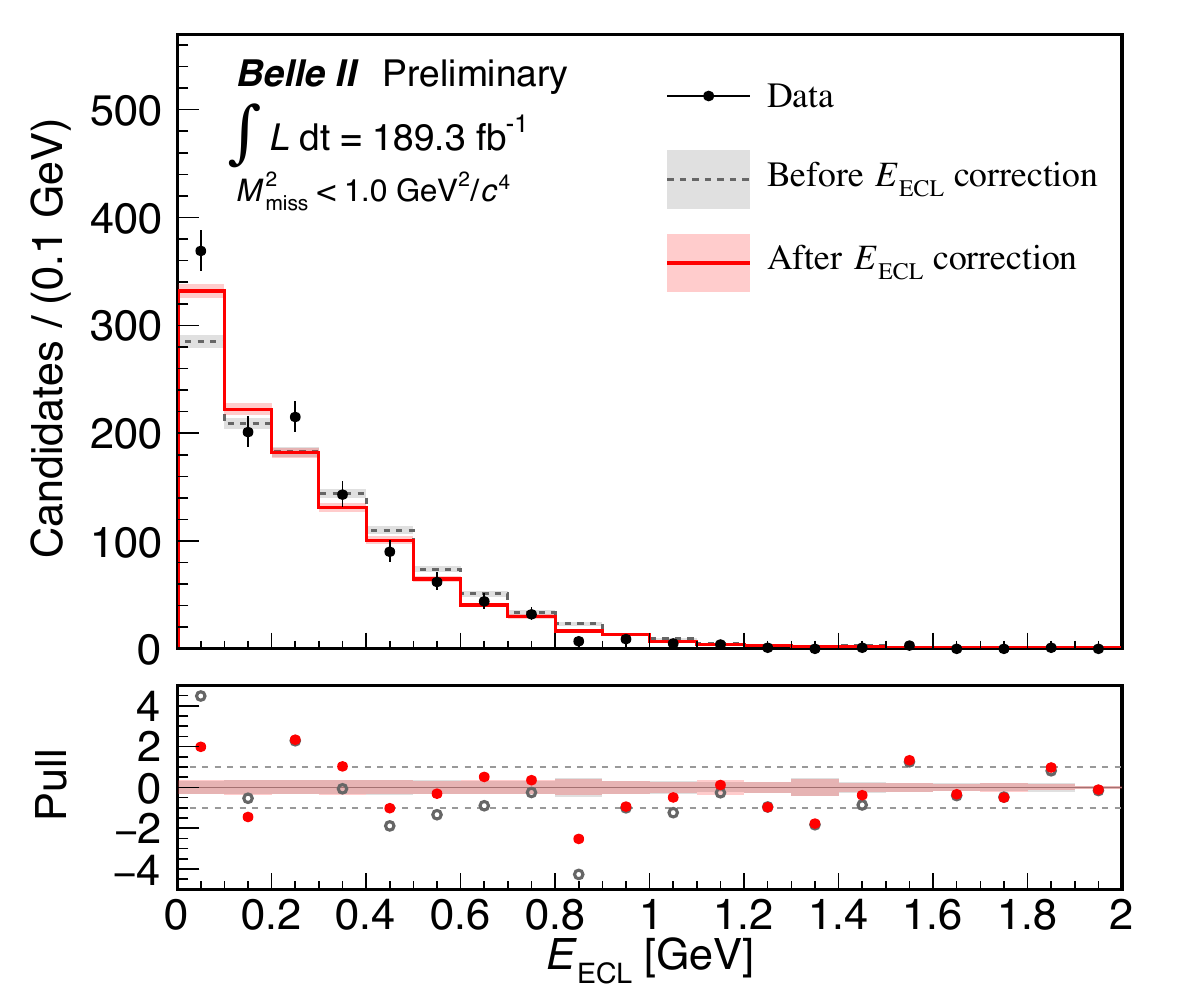}
    \caption{
        Comparison between data and simulation of the $E_{\mathrm{ECL}}$ distributions for ${D^{*}}^{+} \rightarrow D^{0}\pi^{+}$ at $M_{\mathrm{miss}}^{2} < 1.0~\mathrm{GeV}^{2}/c^{4}$. The black points with error bars are the data. The red solid (gray dashed) histogram is from the simulation after (before) the $E_{\mathrm{ECL}}$ correction where the nominal energy shift at $15~\mathrm{MeV}$ is applied. The bottom panel presents pull values before (gray open circles) and after the correction (red closed circles). The rectangular-shaded regions on the histograms and in the pull plot correspond to statistical uncertainties of the simulation.
    }
    \label{fig:EeclCorr}
\end{figure}

As discussed above, $E_{\mathrm{ECL}}$ is the sum of the energies of all neutral clusters in the ECL that are not used in the $\Upsilon$(4S) reconstruction. To validate the shape of $E_{\mathrm{ECL}}$ in the simulation, the normalization mode, $\overline{B} \rightarrow D^{(*)} \ell^{-} \overline{\nu}_{\ell}$, is used as a control sample. The $E_{\mathrm{ECL}}$ distribution has the same properties as in the signal mode and is dominated by background photons. The potential additional contribution from electron bremsstrahlung photons in the $B_{\mathrm{sig}}$ decay is determined to be negligible by comparing the $E_{\mathrm{ECL}}$ distributions in the electron and muon channels. The shape of the $E_{\mathrm{ECL}}$ distribution differs between data and simulation in the region $M_{\mathrm{miss}}^{2} < 1.0~\mathrm{GeV}^{2}/c^{4}$. The main sources of this discrepancy are related to the modeling of neutral clusters resulting from beam-induced backgrounds and the interaction of hadronic particles with detector material (hadronic split-off showers). To address this issue, we replace the simulated beam background contribution with simulation that includes the observed random trigger distribution of each data collection period, i.e. run-dependent simulation. Furthermore, the modeling of hadronic split-off showers is corrected by subtracting $15~\mathrm{MeV}$ from the energy of each neutral cluster produced by a hadronic split-off shower. This energy shift is determined by scanning the range from zero to $30~\mathrm{MeV}$ and finding the energy shift value that minimizes the $\chi^{2}$ between the data and simulated $E_{\mathrm{ECL}}$ distributions. The uncertainty due to the energy shift of split-off showers is then determined to be ${}_{-7}^{+9}~\mathrm{MeV}$, which covers the range $\Delta \chi^{2} < 4$. The difference before and after these $E_{\mathrm{ECL}}$ corrections is illustrated in Figure~\ref{fig:EeclCorr}. Data-simulation comparisons with the aforementioned improvements show consistency among the three $D^{*}$ modes.   

\subsection{$M_{\mathrm{miss}}^{2}$ resolution}

The $M_{\mathrm{miss}}^{2}$ resolution in data is found to be $17\%$ and $15\%$ worse than the resolution in simulation for the $B^{0}$ and $B^{+}$ modes, respectively. This discrepancy primarily arises from the beam energy spread in the data, whereas the beam energies in simulation are set to their nominal values. To address this, the $M_{\mathrm{miss}}^{2}$ distribution is smeared in the simulation by $0.061~\mathrm{GeV}^{2}/c^{4}$ for $B^{0}$ modes and $0.060~\mathrm{GeV}^{2}/c^{4}$ for $B^{+}$ modes. The smearing factor, denoted as $\Delta\sigma_{M_\mathrm{miss}^{2}}$, is determined as
\begin{align}
    \Delta\sigma_{M_\mathrm{miss}^{2}} = \sqrt{\left(\sigma_{M_\mathrm{miss}^{2}}^{\mathrm{data}}\right)^{2} - \left(\sigma_{M_\mathrm{miss}^{2}}^{\mathrm{MC}}\right)^{2}},
    \label{eq:Mmiss2SmearFactor}
\end{align}
where $\sigma_{M_\mathrm{miss}^{2}}^{\mathrm{data}}$ and $\sigma_{M_\mathrm{miss}^{2}}^{\mathrm{MC}}$ are the widths of the $M_{\mathrm{miss}}^{2}$ peak in data and simulation, respectively. These peak widths are obtained from fits in the $|M_{\mathrm{miss}}^{2}| < 0.5~\mathrm{GeV}^{2}/c^{4}$ region.

\section{Sample Composition and Fit}
\label{sec:fit}

\begin{figure*}
    \centering
    \begin{tabular}{ccc}
        \includegraphics[width=0.33\linewidth]{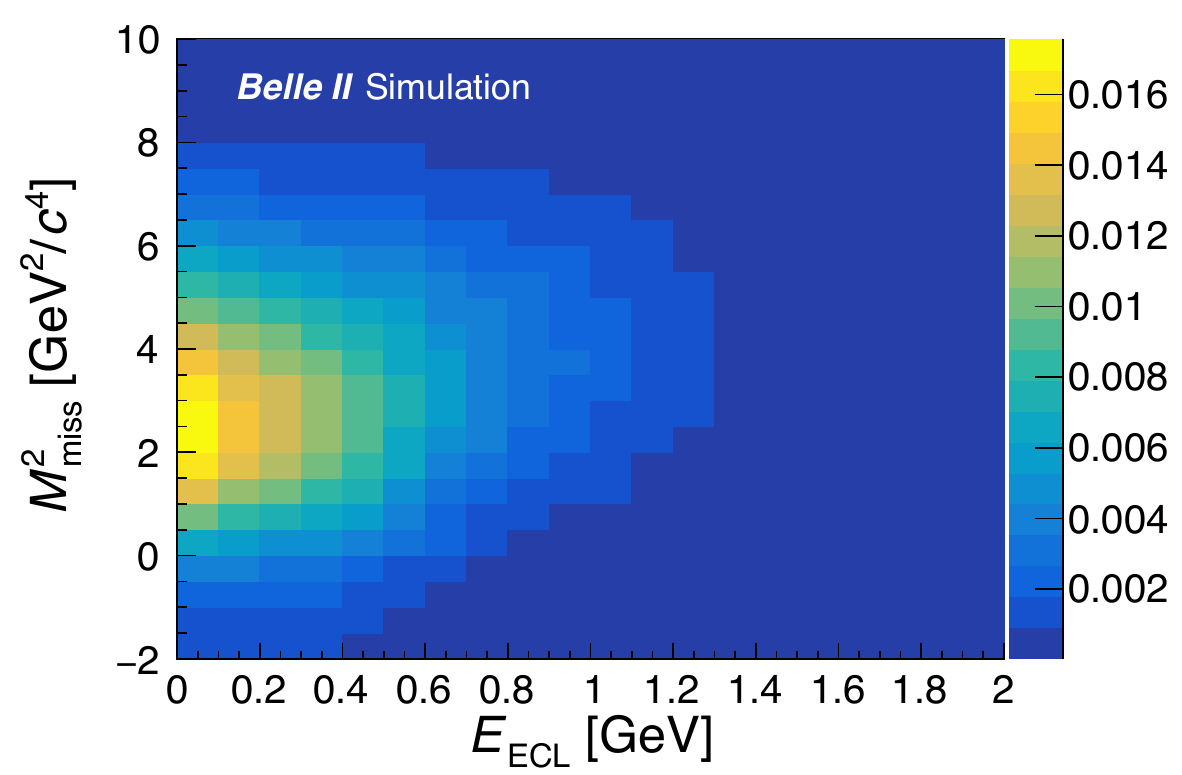} &
        \includegraphics[width=0.33\linewidth]{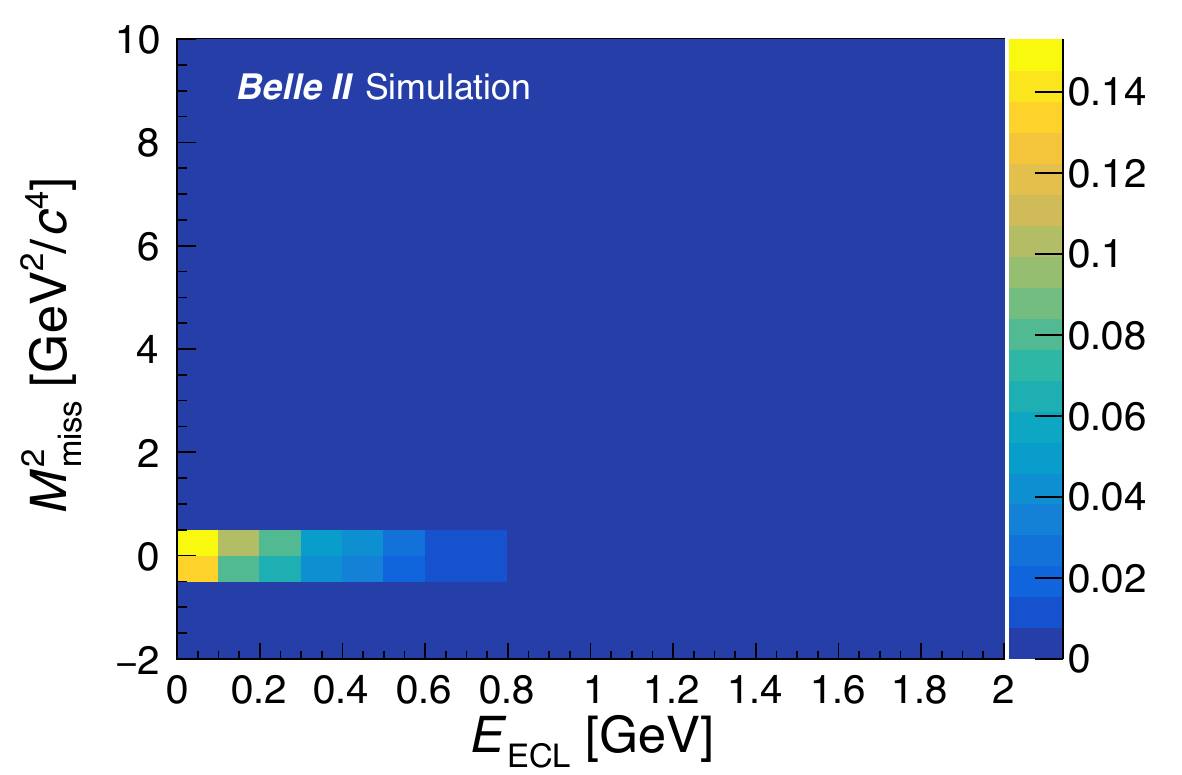} &
        \includegraphics[width=0.33\linewidth]{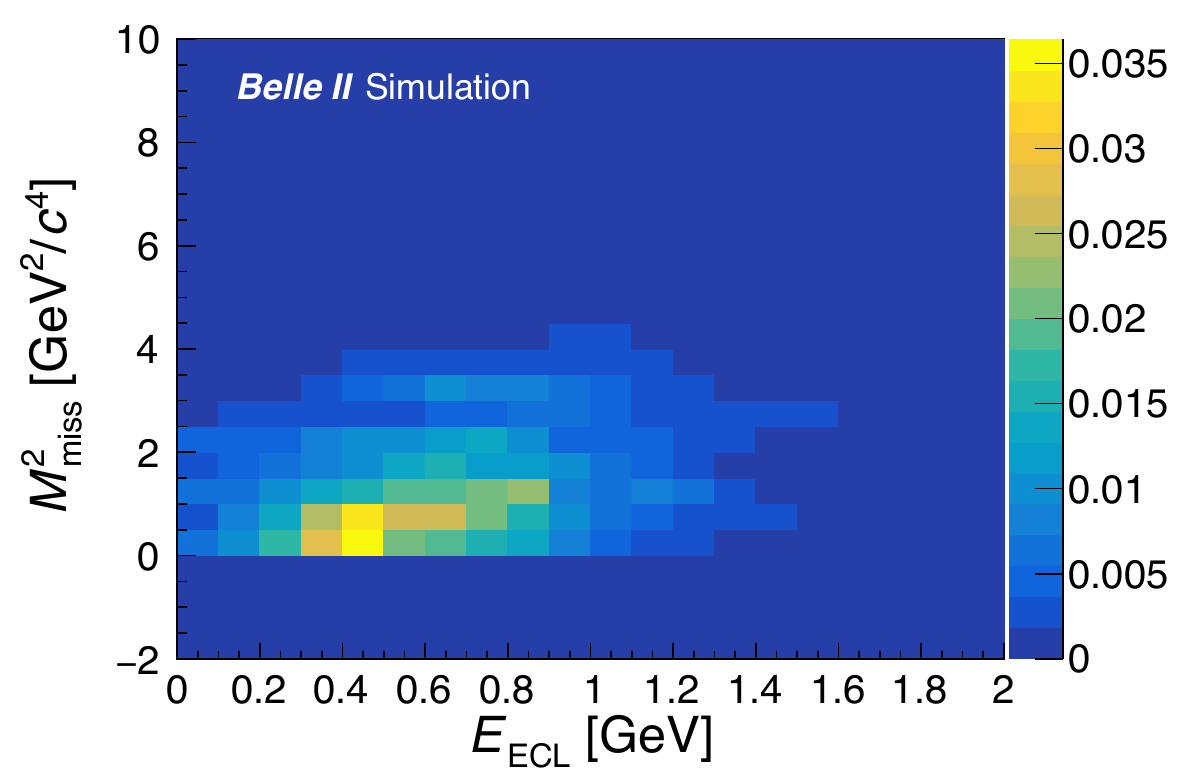}
    \end{tabular}
    \caption{
        Two-dimensional PDFs of $E_{\mathrm{ECL}}$ and $M_{\mathrm{miss}}^{2}$ from $\overline{B} \rightarrow D^{*}\tau^{-}\overline{\nu}_{\tau}$ (left), $\overline{B} \rightarrow D^{*}\ell^{-}\overline{\nu}_{\ell}$ (middle), and $\overline{B} \rightarrow D^{**}\ell^{-}\overline{\nu}_{\ell}$ (right) in the ${D^{*}}^{+} \rightarrow D^{0}\pi^{+}$ mode. The normalization of the intensity scale is arbitrary. 
    }
    \label{fig:pdf}
\end{figure*}

We extract $R(D^{*})$ from an extended two-dimensional likelihood fit to the binned $E_{\mathrm{ECL}}$ and $M_{\mathrm{miss}}^{2}$ distributions. Two-dimensional probability density functions (PDFs) are constructed for each of the three $D^{*}$ decay modes: ${D^{*}}^{+}\rightarrow D^{0}\pi^{+}$, ${D^{*}}^{+}\rightarrow D^{+}\pi^{0}$, and ${D^{*}}^{0}\rightarrow D^{0}\pi^{0}$. A simultaneous fit is performed using the three $D^{*}$ decays. As shown in Figure~\ref{fig:pdf}, the contributions from $\overline{B}\rightarrow D^{*}\tau^{-} \overline{\nu}_{\tau}$, $\overline{B} \rightarrow D^{*} \ell^{-} \overline{\nu}_{\ell}$, and background populate different regions of the $E_{\mathrm{ECL}}$-$M_{\mathrm{miss}}^{2}$ plane.  

For each $D^{*}$ mode, $R(D^{*})$ is calculated using
\begin{align}
    R(D^{*}) &= \frac{\mathcal{B}\left(\overline{B} \rightarrow D^{*}\tau^{-}\nu\right)}{\mathcal{B}\left(\overline{B} \rightarrow D^{*}\ell^{-}\nu\right)} \nonumber \\
    &= \frac{N_{D^{*}\tau\nu}}{(N_{D^{*}\ell\nu}/2)} \frac{\varepsilon_{D^{*}\ell\nu}}{\varepsilon_{D^{*}\tau\nu}},
    \label{eq:fitRDst}
\end{align}
where $N_{D^{*}\tau(\ell)\nu}$ is the observed number of $D^{*}\tau(\ell)\nu$ decays in the data and $\varepsilon_{D^{*}\tau(\ell)\nu}$ is the reconstruction efficiency for correctly reconstructed $B \rightarrow D^{*}\tau(\ell)\nu$ decays. The factor of two in the denominator averages the summed yield from the two modes with light leptons. We assume isospin symmetry for charged and neutral $B$ meson decays and set  $R(D^{*}) = R(D^{*0}) = R(D^{*+})$. Here, the reconstruction efficiencies are defined as
\begin{equation}
    \varepsilon_{D^{*}\tau\nu(D^{*}\ell\nu)} = \frac{N_{D^{*}\tau\nu(D^{*}\ell\nu)}^{\mathrm{rec}}}{N_{D^{*}\tau\nu(D^{*}\ell\nu)}^{\mathrm{gen}}},
\end{equation}
where $N_{D^{*}\tau(\ell)\nu}^{\mathrm{rec}}$ and $N_{D^{*}\tau(\ell)\nu}^{\mathrm{gen}}$ are the numbers of correctly reconstructed and generated $D^{*}\tau(\ell)\nu$ decays in the simulation, respectively.

Some of the fit parameters are unconstrained while others are subjected to Gaussian constraints. We define four event categories in the fit and additionally divide the background events with a correctly reconstructed $D^{*}$ candidate into five subcategories.
The yields of each category or subcategory are parametrized for each $D^{*}$ mode as follows.

\begin{enumerate}
    \item Signal events. \\
    The yield $N_{D^{*}\tau\nu}$ is parametrized by 
    \begin{equation}
        N_{D^{*}\tau\nu} = R(D^{*}) \frac{N_{D^{*}\ell\nu}}{2} \frac{\varepsilon_{D^{*}\tau\nu}}{\varepsilon_{D^{*}\ell\nu}}.
        \label{eq:NDsttaunu}
    \end{equation}
    Here $R(D^{*})$ is unconstrained in the fit, and the reconstruction efficiencies for the signal and the normalization modes are nuisance parameters constrained to the simulated value in each $D^{*}$ mode.
    
    $\overline{B} \rightarrow D^{*}\tau^{-}\overline{\nu}_{\tau}$ decays accompanying a fake lepton candidate from the $\tau$ decay that passes the lepton ID requirement are also treated as signal. Their yield $N_{D^{*}\tau\nu,\ell\text{-}\mathrm{misID}}$ is fixed relative to the yield $N_{D^{*}\tau\nu}$.
    
    \item Normalization events. \\
    The yield $N_{D^{*}\ell\nu}$ depends on the $\overline{B} \rightarrow D^{*}\ell^{-}\overline{\nu}_{\ell}$ branching fractions; the reconstruction efficiency $\varepsilon_{D^{*}\ell\nu}$; $N_{B\overline{B}}$; and the $\Upsilon(\mathrm{4S})$ branching fractions into neutral, $f_{00} = 0.486 \pm 0.012$~\cite{Belle:f00}, or charged $B$ mesons, $f_{+-} = 1 - f_{00}$, \vspace{1.0ex}

    $N_{D^{*}\ell\nu}$
    \begin{numcases}{\hspace{\leftmargin}=}
        2\mathcal{B}(\overline{B}{}^{0} \rightarrow {D^{*}}^{+}\ell^{-}\overline{\nu}_{\ell}) 2N_{B\overline{B}} f_{00} \varepsilon_{D^{*}\ell\nu}, 
        \label{eq:NB0Dstlnu} \\
        2\mathcal{B}(B^{-} \rightarrow {D^{*}}^{0}\ell^{-}\overline{\nu}_{\ell}) 2N_{B\overline{B}} f_{+-} \varepsilon_{D^{*}\ell\nu}.
        \label{eq:NBplusDstlnu}
    \end{numcases}
    The efficiency $\varepsilon_{D^{*}\ell\nu}$ is a parameter specific to each $D^{*}$ modes.
    The branching fractions $\mathcal{B}(\overline{B} \rightarrow {D^{*}}\ell^{-}\overline{\nu}_{\ell})$ are unconstrained in the fit, while $f_{00}$, $N_{B\overline{B}}$, and $\varepsilon_{D^{*}\ell\nu}$ are constrained nuisance parameters. 
    
    \item Background events with a correctly reconstructed $D^{*}$ candidate. \\
    The $\overline{B} \rightarrow D^{**}\ell^{-}\overline{\nu}$ yield $N_{D^{**}\ell\nu}$ is unconstrained in the fit as the individual branching fractions in this component are poorly constrained by existing measurements, while the hadronic $B$ decay yield ($N_{\mathrm{Had}B}$), ${B}^{0}\leftrightarrow B^{+}$ cross feed yields of semileptonic decays ($N_{B\mathrm{CF}}$), continuum event yield ($N_{q\overline{q}}$), and other background event yield ($N_{\mathrm{other}}$) are fixed to the values from simulation. The $\overline{B} \rightarrow D^{**}\ell^{-}\overline{\nu}$ category includes $\overline{B} \rightarrow D^{**}\ell^{-}\overline{\nu}_{\ell}$, $\overline{B} \rightarrow D_{\mathrm{gap}}^{**}\ell^{-}\overline{\nu}_{\ell}$, and $\overline{B} \rightarrow D^{**}\tau^{-}\overline{\nu}_{\tau}$ events. The ``other'' background category contains $\overline{B} \rightarrow D^{*}\tau^{-}\overline{\nu}_{\tau}$ and $\overline{B} \rightarrow D^{*}\ell^{-}\overline{\nu}_{\ell}$ events where daughter particles are misassigned between the signal-side and tag-side $B$ mesons.
    
    \item Background events with a fake $D^{*}$ candidate. \\
    The yield $N_{\mathrm{Fake}D^{*}}$ is estimated by the fit with a constraint given by the calibration factor determined in the $\Delta M_{D^{*}}$ fits. 
\end{enumerate}

The treatment of fit parameters is summarized in Table~\ref{tab:fitPars}. In the fits to all $D^{*}$ modes, $R(D^{*})$ is a shared parameter. The value of $\mathcal{B}(\overline{B}{}^{0} \rightarrow {D^{*}}^{+}\ell^{-}\overline{\nu}_{\ell})$ is common to the fit categories of ${D^{*}}^{+} \rightarrow D^{0}\pi^{+}$ and ${D^{*}}^{+} \rightarrow D^{+}\pi^{0}$. The other parameters are determined independently in each $D^{*}$ mode. In total, six parameters are unconstrained in the fit as shown in Table~\ref{tab:fitPars}, while 11 nuisance parameters are constrained in the fit, namely $\varepsilon_{D^{*}\tau\nu}$~(3), $\varepsilon_{D^{*}\ell\nu}$~(3), $N_{\mathrm{Fake}D^{*}}$~(3), $f_{00}$~(1), and $N_{B\overline{B}}$~(1). The values of the fixed parameters are also listed in Table~\ref{tab:fixedParValues}.

\begin{table*}
	\centering
	\caption{
        Summary of fit-parameter configuration. The index $i$ designates the fit category for the three $D^{*}$ decays. The total number of parameters in the fit is indicated in parentheses.
    }
	\begin{tabular}{llcccccc}
		\toprule
        \toprule
		\multicolumn{2}{c}{\multirow{2}{*}{PDF component}} & \multicolumn{6}{c}{Parameter} \\
		\cmidrule{3-8}
		& & \multicolumn{2}{c}{Unconstrained} & & \multicolumn{2}{c}{Fixed} \\ 
		\midrule
        \multicolumn{2}{l}{Signal events} & $R(D^{*})$ & (1) &  &  & $N_{D^{*}\tau\nu,\ell\text{-}\mathrm{misID}}^{i}$/$N_{D^{*}\tau\nu}^{i}$ & (3) \\
        \multicolumn{2}{l}{\multirow{2}{*}{Normalization events}} & $\mathcal{B}\left({\overline{B}{}^{0} \rightarrow {D^{*}}^{+}\ell^{-}\overline{\nu}_{\ell}}\right)$ & \multirow{2}{*}{(2)} & \multirow{2}{*} & \multirow{2}{*}{} & \multirow{2}{*}{---} & \\
        & & $\mathcal{B}\left({B^{-} \rightarrow {{D}^{*}}^{0}\ell^{-}\overline{\nu}_{\ell}}\right)$ & &  & & \\
        \midrule
        \multicolumn{2}{l}{Background events with a correct $D^{*}$} & & \\
        \cmidrule{2-2}
        \hspace{3mm} & $\overline{B} \rightarrow D^{**}\ell^{-}\overline{\nu}$ & $N_{D^{**}\ell\nu}^{i}$ & (3) &  & & --- & \\
        & Hadronic $B$ decay & --- & &  & & $N_{\mathrm{Had}B}^{i}$ & (3) \\
        & ${B}^{0}\leftrightarrow B^{+}$ cross feed & --- & & & & $N_{\mathrm{BCF}}^{i}$ & (3) \\
        & Continuum events & --- & & & & $N_{q\overline{q}}^{i}$ & (3) \\
        & Other background candidates & --- & & & & $N_{\mathrm{other}}^{i}$ & (3) \\
        \cmidrule{2-2}
        \multicolumn{2}{l}{Background events with a fake $D^{*}$} & --- & & & & --- & \\
        \midrule
        \multicolumn{2}{l}{Total} & & (6) & & & & (15) \\
        \bottomrule
        \bottomrule
	\end{tabular}
	\label{tab:fitPars}
\end{table*}

\begin{table*}
    \centering
    \caption{
        Values of the parameters fixed in the fit for $R(D^{*})$. The index $i$ designates the fit category for the three $D^{*}$ decays. 
    }
    \begin{tabular}{lccc}
        \toprule
        \toprule
        \multirow{2}{*}{Parameter} & \multicolumn{3}{c}{Constrained value} \\
        \cmidrule{2-4}
        & ${D^{*}}^{+} \rightarrow D^{0}\pi^{+}$ & ${D^{*}}^{+} \rightarrow D^{+}\pi^{0}$ & ${D^{*}}^{0} \rightarrow D^{0}\pi^{0}$ \\
        \midrule
        $N_{D^{*}\tau\nu,\ell\text{-}\mathrm{misID}}^{i}$/$N_{D^{*}\tau\nu}^{i}$~[$$\%$$] & $6.6$ & $6.6$ & $6.9$ \\
        $N_{\mathrm{Had}B}^{i}$ & $38.8$ & $4.5$ & $21.2$ \\
        $N_{\mathrm{BCF}}^{i}$ & $8.8$ & $1.6$ & $4.5$ \\
        $N_{q\overline{q}}^{i}$ & $1.5$ & $0.0$ & $0.9$ \\
        $N_{\mathrm{other}}^{i}$ & $2.2$ & $0.3$ & $1.5$ \\
        \bottomrule
        \bottomrule
    \end{tabular}
    \label{tab:fixedParValues}
\end{table*}

The PDFs used in the fit are constructed from template distributions based on simulation. The presence of low-population bins in the templates, due to the limited size of simulation samples, introduces a potential bias in the results. An adaptive kernel density estimation (KDE)~\cite{KDE} method is used to smooth the PDF across the bins, and the KDE is applied to all categories except for the normalization events. The KDE densities for different categories of templates are optimized separately. 

The bias of the $R(D^{*})$ estimator is checked for true $R(D^{*})$ values within the range $0.10$--$0.60$. We determine a correction function that relates the fitted value to the true value, of the form
\begin{align}
    R(D^{*})_{\mathrm{true}} = 1.008 \times R(D^{*}) - 0.003,
    \label{eq:fitterLinearity}
\end{align}
where $R(D^{*})_{\mathrm{true}}$ is the true value of $R(D^{*})$. The uncertainty associated with this linear function is found to be at most $0.1\%$ of $R(D^{*})_{\mathrm{true}}$ within the tested range. The observed bias determined from this function is included as a source of systematic uncertainty.

\section{Results}
\label{sec:result}

\begin{figure*}
    \centering
    \subfloat[][]{
        \includegraphics[width=0.33\linewidth]{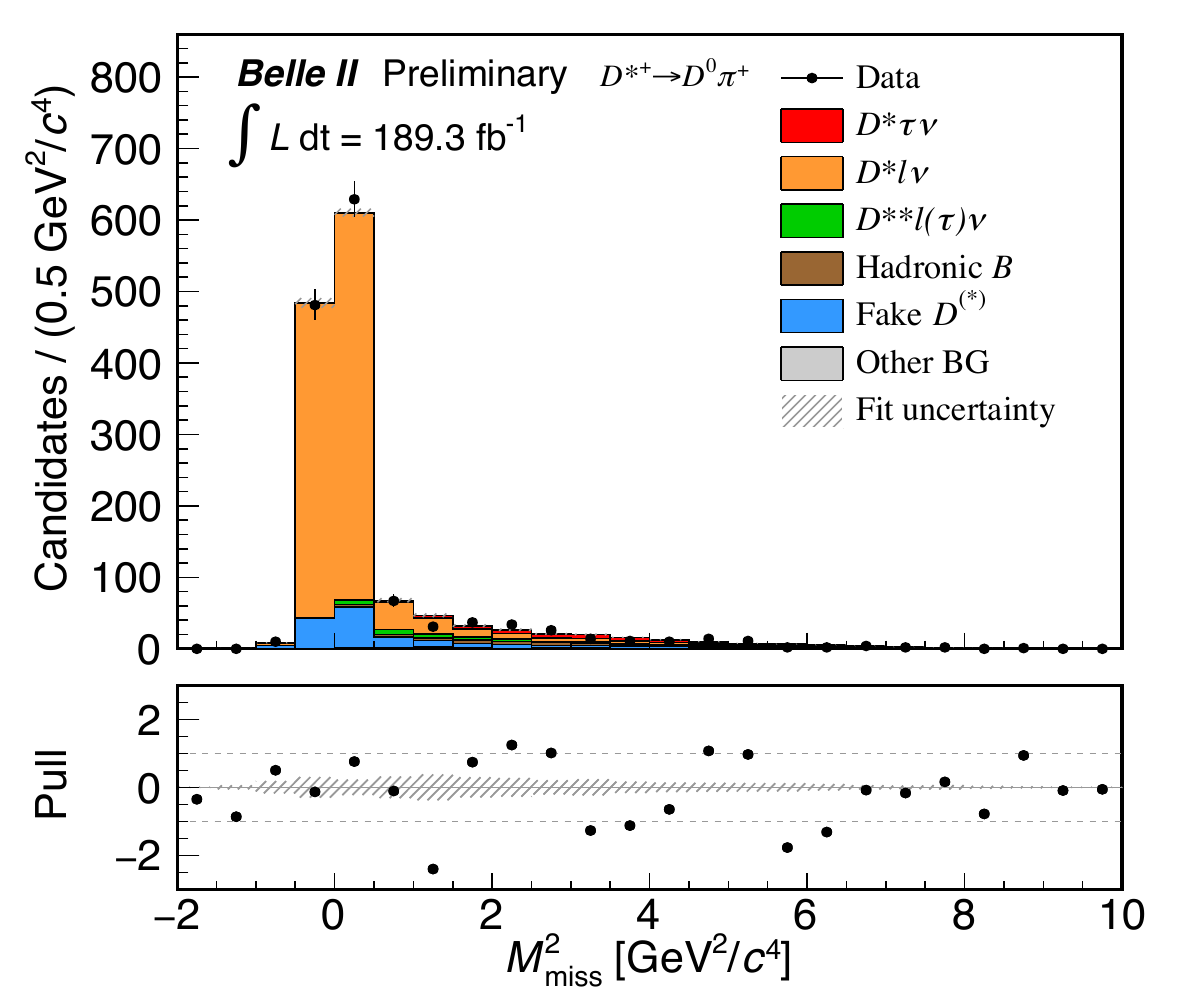}
        \label{fig:Mmiss2Xc100Postfit}
    }
    \subfloat[][]{
        \includegraphics[width=0.33\linewidth]{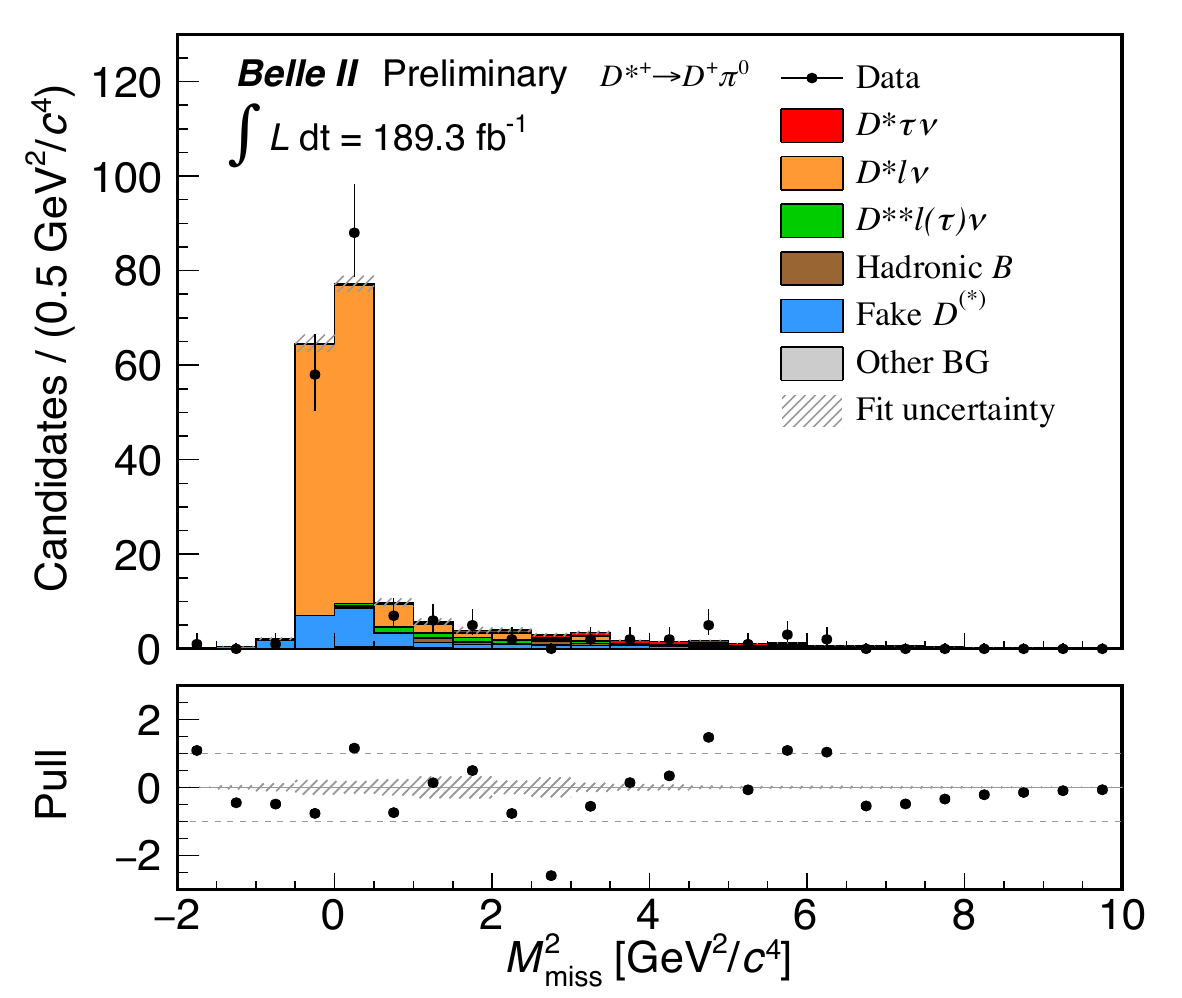}
        \label{fig:Mmiss2Xc200Postfit}
    }
    \subfloat[][]{
        \includegraphics[width=0.33\linewidth]{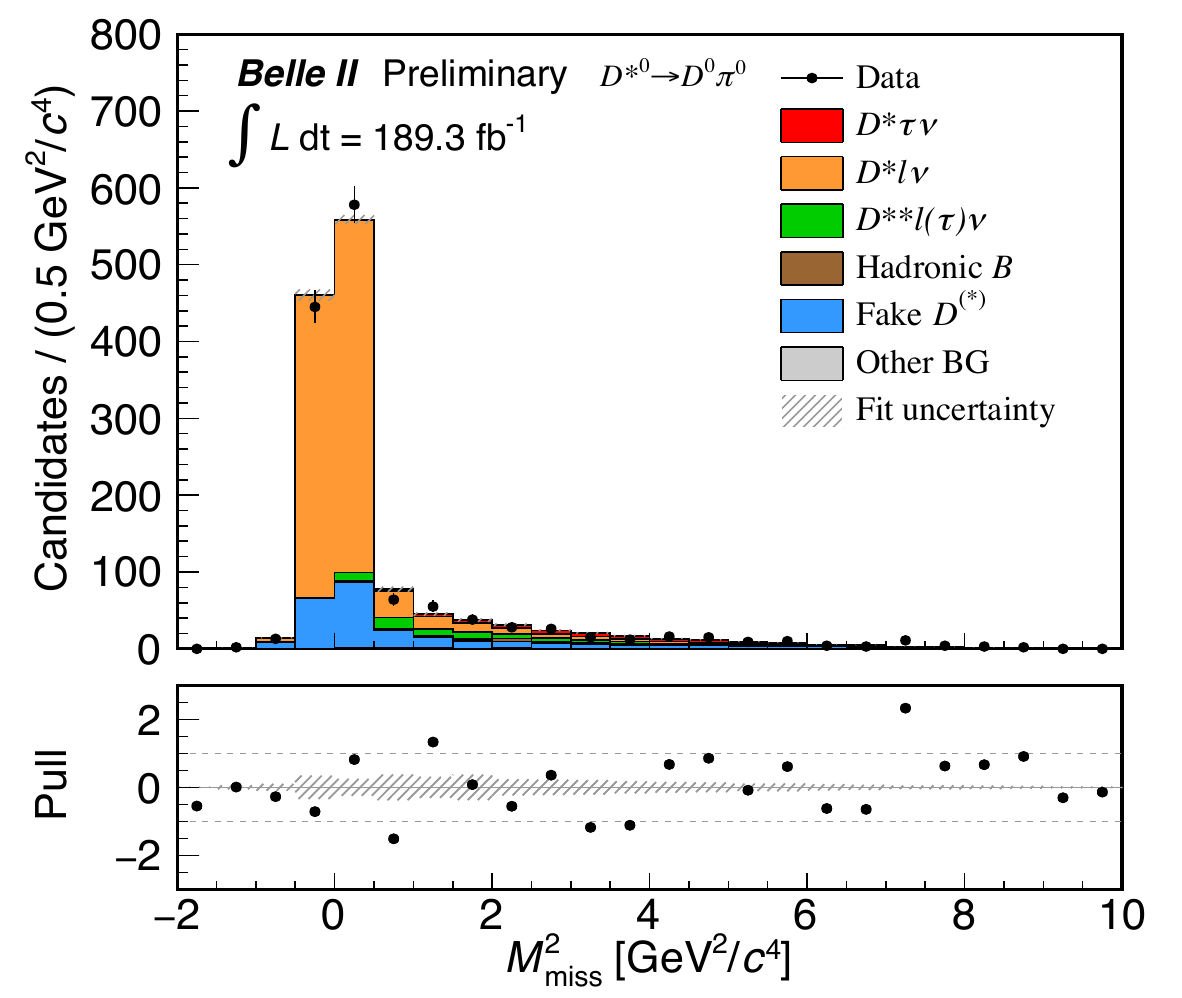}
        \label{fig:Mmiss2Xc300Postfit}
    } \\
    \subfloat[][]{
        \includegraphics[width=0.33\linewidth]{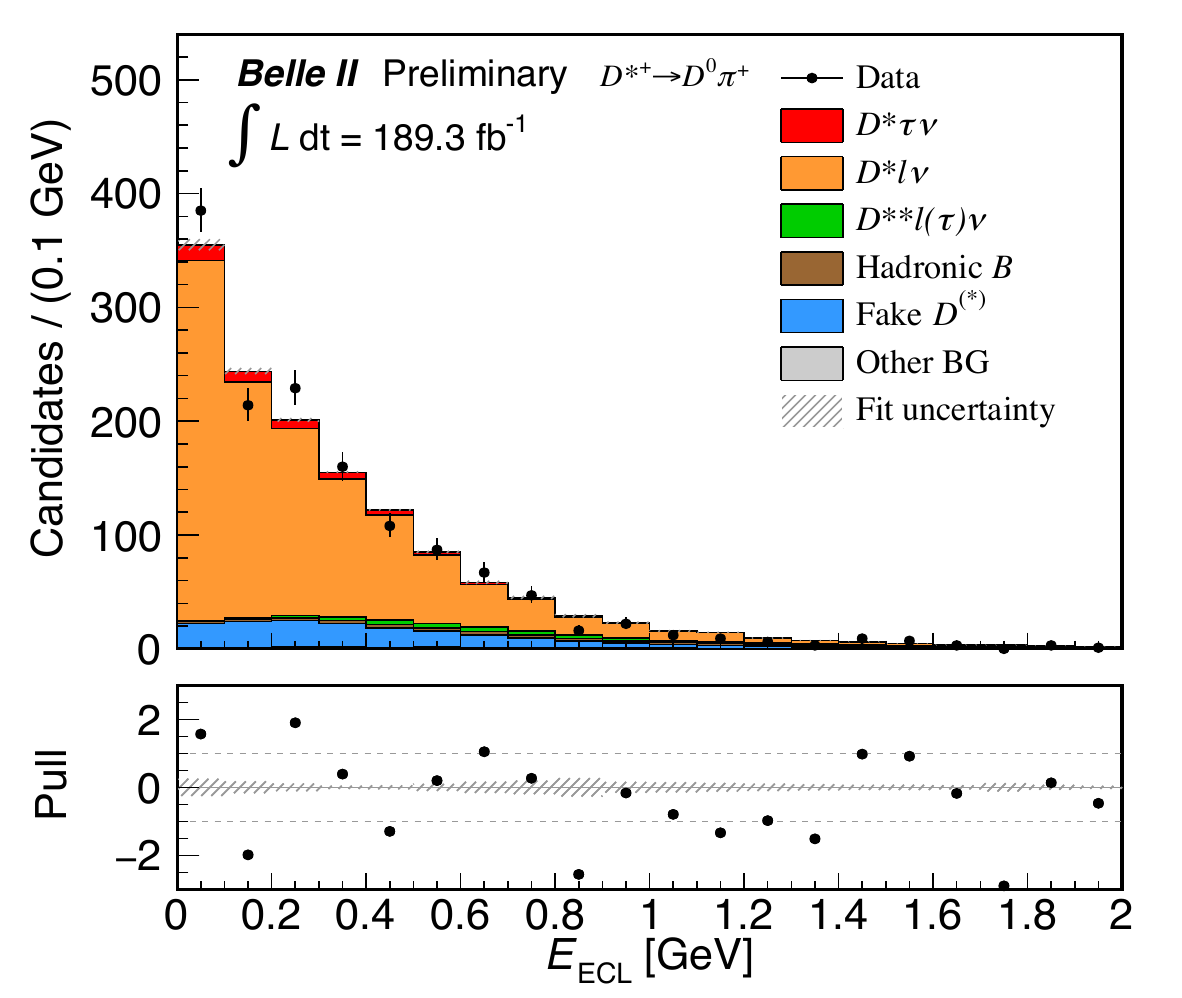}
        \label{fig:EeclXc100Postfit}
    }
    \subfloat[][]{
        \includegraphics[width=0.33\linewidth]{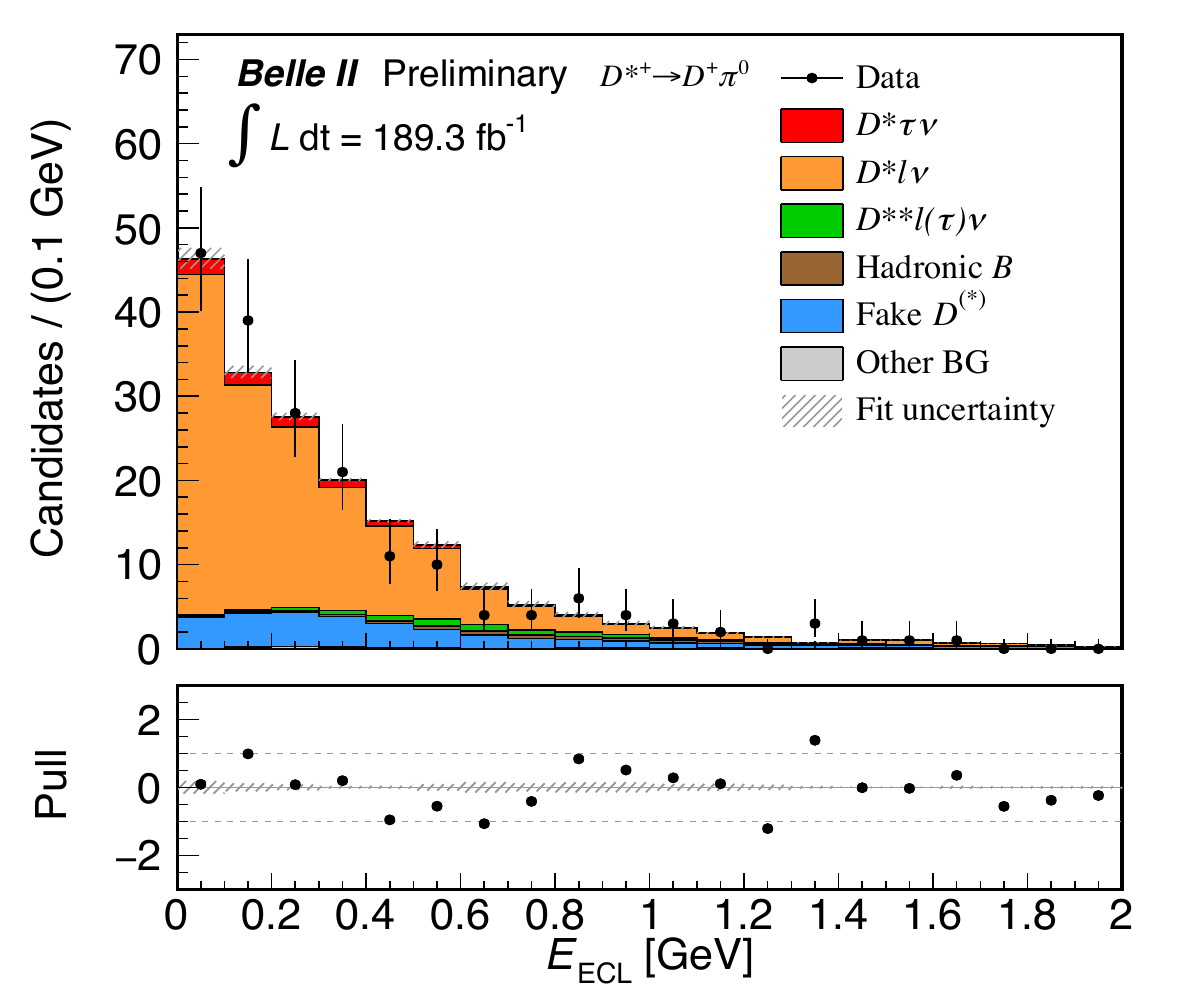}
       \label{fig:EeclXc200Postfit}
    }
    \subfloat[][]{
        \includegraphics[width=0.33\linewidth]{Figures/2023-12-29_P12HadFEI_realDataFitResult_EeclMinC2TDist20Mmiss2_Xc100-300_00000_6.pdf}
        \label{fig:EeclXc300Postfit}
    } \\
    \caption{
        Distributions of \protect\subref{fig:Mmiss2Xc100Postfit}--\protect\subref{fig:Mmiss2Xc300Postfit}~$M_{\mathrm{miss}}^{2}$ and \protect\subref{fig:EeclXc100Postfit}--\protect\subref{fig:EeclXc300Postfit}~$E_{\mathrm{ECL}}$ in the entire region for the 
        ${D^{*}}^{+} \rightarrow D^{0}\pi^{+}$ (left), 
        ${D^{*}}^{+} \rightarrow D^{+}\pi^{0}$ (middle), and 
        ${D^{*}}^{0} \rightarrow D^{0}\pi^{0}$ (right) modes, with fit projections overlaid. The bottom panel presents pull values from fit results. The rectangular-shaded regions on the histograms and in the pull plot correspond to statistical uncertainties in the fit.
    }
    \label{fig:Postfit}
\end{figure*}

\begin{figure*}
    \centering
    \subfloat[][]{
        \includegraphics[width=0.33\linewidth]{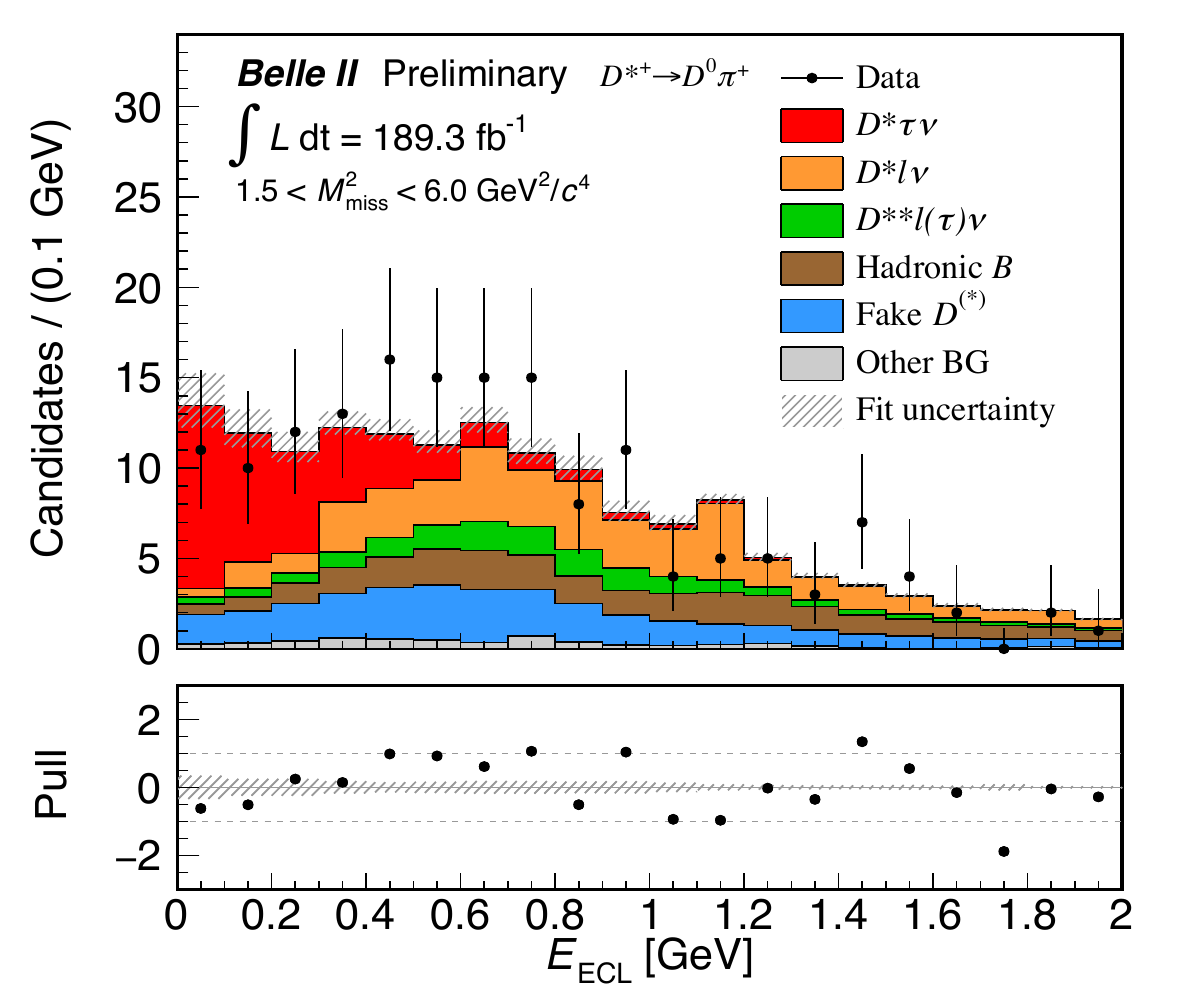}
        \label{fig:EeclXc100PostfitMiddleMmiss2}
    }
    \subfloat[][]{
        \includegraphics[width=0.33\linewidth]{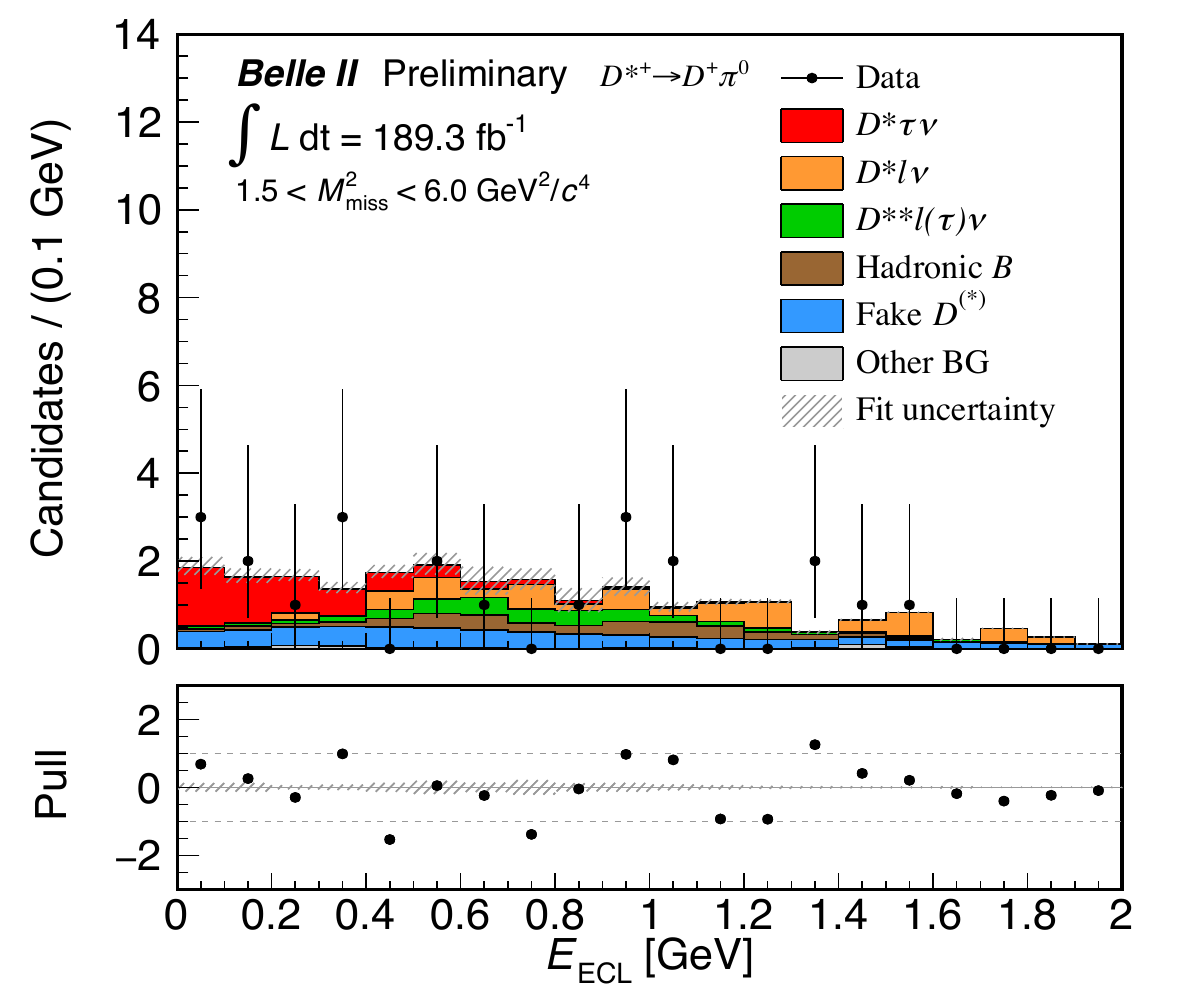}
        \label{fig:EeclXc200PostfitMiddleMmiss2}
    }
    \subfloat[][]{
        \includegraphics[width=0.33\linewidth]{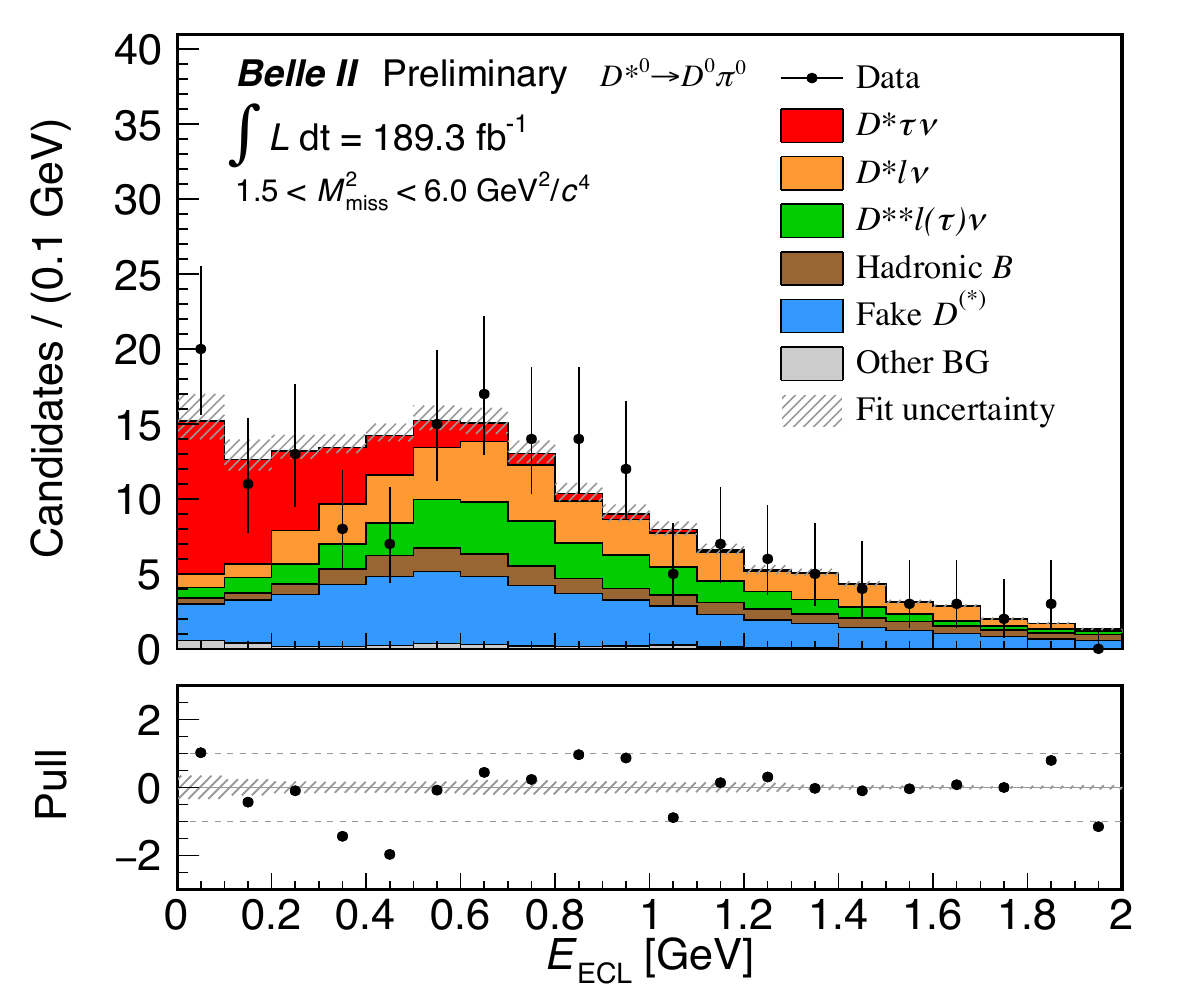}
        \label{fig:EeclXc300PostfitMiddleMmiss2}
    } \\
    \caption{
        Distributions of $E_{\mathrm{ECL}}$ in the signal-enhanced region $1.5 < M_{\mathrm{miss}}^{2} < 6.0~\mathrm{GeV}^{2}/c^{4}$ for the 
        ${D^{*}}^{+} \rightarrow D^{0}\pi^{+}$ (left), 
        ${D^{*}}^{+} \rightarrow D^{+}\pi^{0}$ (middle), and 
        ${D^{*}}^{0} \rightarrow D^{0}\pi^{0}$ (right) modes, with fit projections overlaid. The bottom panel presents pull values from fit results. The rectangular-shaded regions on the histograms and in the pull plot correspond to statistical uncertainties in the fit.
    }
    \label{fig:PostfitSigEnhanced}
\end{figure*}

\begin{figure}
    \centering\includegraphics[width=\linewidth]{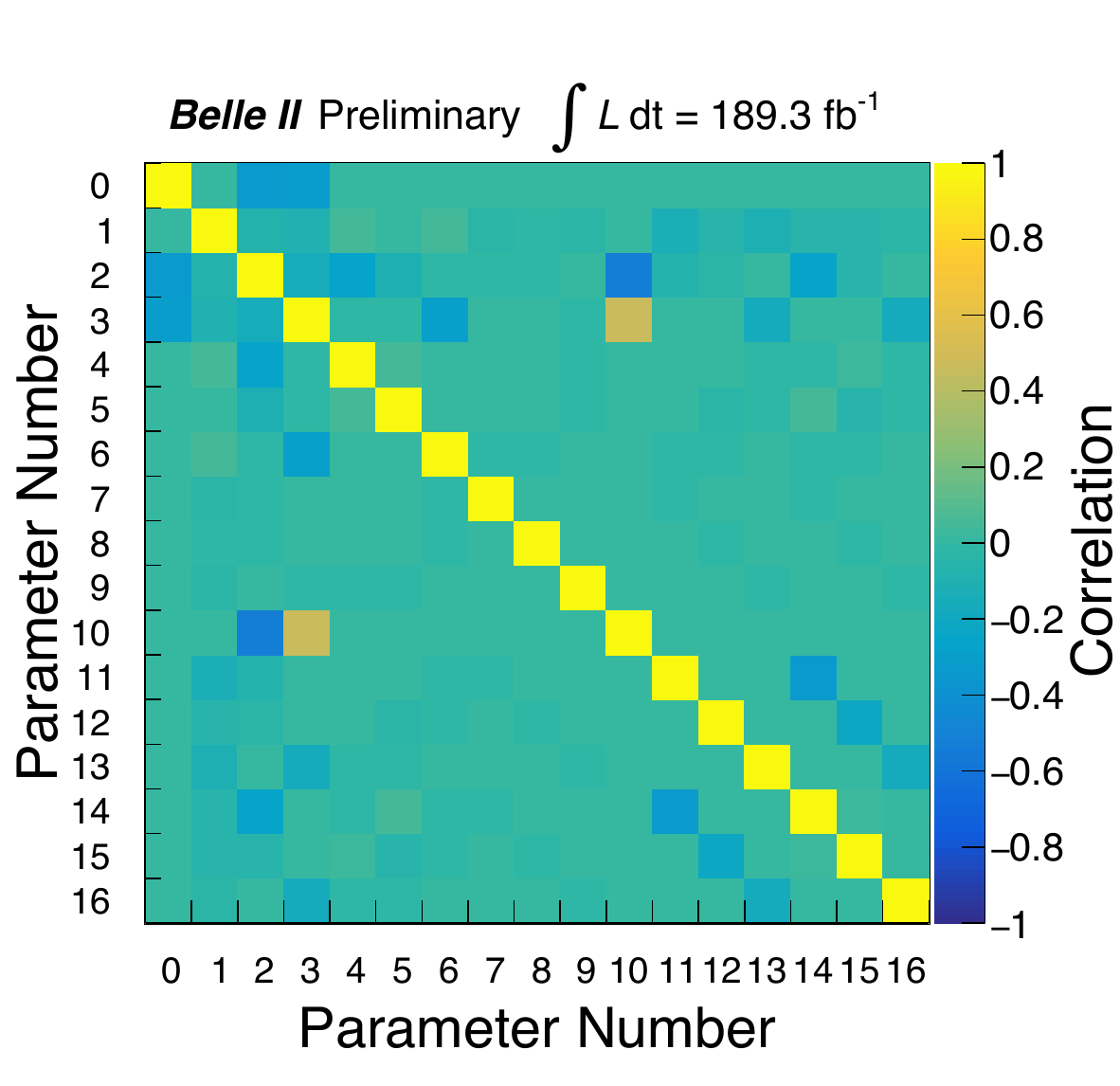}
    \caption{
        Correlation matrix of the fit parameters in the extraction of $R(D^{*})$. The axes are identification numbers of these parameters that refer to Table~\ref{tab:RDstFitParIDs}.
    }
    \label{fig:RDstFitCorrMat}
\end{figure}

The fit to the entire data sample gives
\begin{align}
    R(D^{*}) = 0.262~_{-0.039}^{+0.041}
\end{align}
corresponding to a yield of $108 \pm 16~\overline{B} \rightarrow D^{*}\tau^{-} \overline{\nu}_{\tau}$ events. Here the uncertainties are statistical only. The $p$-value for the goodness of fit is $4.4\%$ according to the $\chi^{2}$ distribution of simplified simulated experiments based on sampling events from the likelihood. Projections of the fit are shown in Figures~\ref{fig:Postfit}~(a)-(c) and (d)-(f) for the $M_{\mathrm{miss}}^{2}$ and $E_{\mathrm{ECL}}$ distributions, respectively. Figures~\ref{fig:PostfitSigEnhanced} show the signal-enhanced ($1.5~\mathrm{GeV}^{2}/c^{4} < M_{\mathrm{miss}}^{2} < 6.0~\mathrm{GeV}^{2}/c^{4}$) fit projection to the $E_{\mathrm{ECL}}$ distribution. Results are given in Table~\ref{tab:finalFitParValues} for the unconstrained parameters and in Table~\ref{tab:finalNuisanceParValues} for the nuisance parameters. Figure~\ref{fig:RDstFitCorrMat} shows the correlation matrix of these unconstrained and nuisance parameters. The matrix is presented with axes of identification numbers of the fit parameters listed in Table~\ref{tab:RDstFitParIDs}. The efficiency ratios $\varepsilon_{D^{*}\tau\nu} / \varepsilon_{D^{*}\ell\nu}$ determined by the fit are $0.336 \pm 0.006$, $0.406 \pm 0.018$, and $0.373 \pm 0.007$ for ${D^{*}}^{+} \rightarrow D^{0}\pi^{+}$, ${D^{*}}^{+} \rightarrow D^{+}\pi^{0}$, and ${D^{*}}^{0} \rightarrow D^{0}\pi^{0}$, respectively. The results for the branching fractions are
\begin{align}
    \mathcal{B}(\overline{B}{}^{0} \rightarrow {D^{*}}^{+}\ell^{-}\overline{\nu}_{\ell}) = \left(5.27~_{-0.24}^{+0.25}\right)\%, \\
    \mathcal{B}(B^{-} \rightarrow {D^{*}}^{0}\ell^{-}\overline{\nu}_{\ell}) = \left(5.50~_{-0.27}^{+0.28}\right)\%.
\end{align}
These branching fractions agree with the current world averages~\cite{pdg:2022}. The yields of $\overline{B} \rightarrow D^{**}\ell^{-} \overline{\nu}_{\ell}$ events are also consistent with the expectations within two statistical standard deviations ($\sigma$) for all $D^{*}$ modes. Taking into account the $30\%$--$40\%$ uncertainties due to the branching fractions of $\overline{B} \rightarrow D^{**}\ell^{-} \overline{\nu}_{\ell}$ decays, the discrepancies from the expectations decrease to $0.5\sigma$--$0.9\sigma$. The yields of the signal and normalization events are calculated using the fitted values of the unconstrained and nuisance parameters of the final fit and summarized in Table~\ref{tab:fittedYields}.

\begin{table*}
    \centering
    \caption{
        Observed (expected) values of the parameters unconstrained in the fit. The expected values are isospin-averaged branching fractions or simulated yields. The index $i$ designates the fit category for the three $D^{*}$ decays. Only statistical uncertainties are given.
    }
    \begin{tabular}{llrlrlrl}
        \toprule
        \toprule
        Parameter & & \multicolumn{6}{c}{Observed (expected) value} \\
        \midrule
        $R(D^{*})$ & & & & \multicolumn{2}{c}{$0.262~_{-0.039}^{+0.041}$} & & \rule[0mm]{0mm}{2.0mm} \\
        $\mathcal{B}(\overline{B}{}^{0} \rightarrow {D^{*}}^{+}\ell^{-}\bar{\nu}_{\ell})~[\%]$ & & & & $5.27~_{-0.24}^{+0.25}$ & ($5.03 \pm 0.11$) & & \rule[0mm]{0mm}{4.0mm} \\
        $\mathcal{B}(B^{-} \rightarrow {D^{*}}^{0}\ell^{-}\bar{\nu}_{\ell})~[\%]$ & & & & $5.50~_{-0.27}^{+0.28}$ & ($5.41 \pm 0.11$) & & \rule[0mm]{0mm}{4.0mm} \\
        \cmidrule{3-8}
        & & \multicolumn{2}{c}{${D^{*}}^{+} \rightarrow D^{0}\pi^{+}$} & \multicolumn{2}{c}{${D^{*}}^{+} \rightarrow D^{+}\pi^{0}$} & \multicolumn{2}{c}{${D^{*}}^{0} \rightarrow D^{0}\pi^{0}$} \\
        \cmidrule{3-8}
        $N_{D^{**}\ell\nu}^{i}$ & & $34.7_{-18.2}^{+19.2}$ & ($61.6 \pm 2.2$)~~ & $5.8_{-4.7}^{+5.6}$ & ($9.0 \pm 0.9$)~~ & $64.5_{-18.3}^{+19.3}$ & ($46.0 \pm 2.0$)~~ \rule[0mm]{0mm}{4.0mm} \\
        \bottomrule
        \bottomrule
    \end{tabular}
    \label{tab:finalFitParValues}
\end{table*}

\begin{table*}
    \centering
    \caption{
        Observed (expected) values of the nuisance parameters. The central values and uncertainties on the expected parameter values are used in the parameter constraints. The index $i$ designates the fit category for the three $D^{*}$ decays. Only statistical uncertainties are given.
    }
    \begin{tabular}{llrlrlrl}
        \toprule
        \toprule
        Parameter & & \multicolumn{6}{c}{Observed (expected) value} \\
        \midrule
        $f_{00}$ & & & & $0.484 \pm 0.012$ & ($0.484 \pm 0.012$) & & \rule[0mm]{0mm}{2.0mm} \\
        $N_{B\bar{B}}$~[$10^{6}$] & & & & $198.0 \pm 3.0$ & ($198.0 \pm 3.0$) & & \rule[0mm]{0mm}{4.0mm} \\
        \cmidrule{3-8}
        & & \multicolumn{2}{c}{${D^{*}}^{+} \rightarrow D^{0}\pi^{+}$} & \multicolumn{2}{c}{${D^{*}}^{+} \rightarrow D^{+}\pi^{0}$} & \multicolumn{2}{c}{${D^{*}}^{0} \rightarrow D^{0}\pi^{0}$} \\
        \cmidrule{3-8}
        $\varepsilon_{D^{*}\tau\nu}^{i}$ [$10^{-5}$] & & $1.805 \pm 0.018$ & ($1.805 \pm 0.018$)~~ & $0.277 \pm 0.007$ & ($0.277 \pm 0.007$)~~ & $1.565 \pm 0.017$ & ($1.565 \pm 0.017$)~~ \rule[0mm]{0mm}{4.0mm} \\
        $\varepsilon_{D^{*}\ell\nu}^{i}$ [$10^{-5}$] & & $5.368 \pm 0.075$ & ($5.363 \pm 0.075$)~~ & $0.683 \pm 0.025$ & ($0.686 \pm 0.027$)~~ & $4.190_{-0.062}^{+0.063}$ & ($4.192 \pm 0.063$)~~ \rule[0mm]{0mm}{4.0mm} \\
        $N_{\mathrm{Fake}D^{*}}^{i}$ & & $164.7_{-22.6}^{+21.9}$ & ($160.7_{-25.5}^{+23.7}$) & $28.8_{-4.2}^{+4.1}$ & ($27.6_{-4.8}^{+4.4}$) & $258.9_{-12.7}^{+12.7}$ & ($251.4_{-13.6}^{+13.1}$) \rule[0mm]{0mm}{4.0mm} \\
        \bottomrule
        \bottomrule
    \end{tabular}
    \label{tab:finalNuisanceParValues}
\end{table*}

\begin{table}
    \centering
    \caption{
        Identification numbers of the fit parameters used in Figure~\ref{fig:RDstFitCorrMat}.
    }
    \begin{tabular}{cl}
        \toprule
        \toprule
        ~~~Number~~~ & Fit parameter\\
        \midrule
        0 & $N_{B\overline{B}}$ \\
        1 & $R(D^{*})$ \\
        2 & $\mathcal{B}(\overline{B}^{0} \rightarrow {D^{*}}^{+}\ell^{-}\overline{\nu}_{\ell})$~\hspace{0.5cm} \\
        3 & $\mathcal{B}(B^{-} \rightarrow {D^{*}}^{0}\ell^{-}\overline{\nu}_{\ell})$ \\
        4 & $\varepsilon_{D^{*}\ell\nu}^{{D^{*}}^{+} \rightarrow D^{0}\pi^{+}}$ \\
        5 & $\varepsilon_{D^{*}\ell\nu}^{{D^{*}}^{+} \rightarrow D^{+}\pi^{0}}$ \\
        6 & $\varepsilon_{D^{*}\ell\nu}^{{D^{*}}^{0} \rightarrow D^{0}\pi^{0}}$ \\
        7 & $\varepsilon_{D^{*}\tau\nu}^{{D^{*}}^{+} \rightarrow D^{0}\pi^{+}}$ \\
        8 & $\varepsilon_{D^{*}\tau\nu}^{{D^{*}}^{+} \rightarrow D^{+}\pi^{0}}$ \\
        9 & $\varepsilon_{D^{*}\tau\nu}^{{D^{*}}^{0} \rightarrow D^{0}\pi^{0}}$ \\
        10 & $f_{00}$ \\
        11 & $N_{D^{**}\ell\nu}^{{D^{*}}^{+} \rightarrow D^{0}\pi^{+}}$ \\
        12 & $N_{D^{**}\ell\nu}^{{D^{*}}^{+} \rightarrow D^{+}\pi^{0}}$ \\
        13 & $N_{D^{**}\ell\nu}^{{D^{*}}^{0} \rightarrow D^{0}\pi^{0}}$ \\
        14 & $N_{\mathrm{Fake}D^{*}}^{{D^{*}}^{+} \rightarrow D^{0}\pi^{+}}$ \\
        15 & $N_{\mathrm{Fake}D^{*}}^{{D^{*}}^{+} \rightarrow D^{+}\pi^{0}}$ \\
        16 & $N_{\mathrm{Fake}D^{*}}^{{D^{*}}^{0} \rightarrow D^{0}\pi^{0}}$ \\
        \bottomrule
        \bottomrule
    \end{tabular}
    \label{tab:RDstFitParIDs}
\end{table}

\begin{table*}
    \centering
    \caption{
        Observed (expected) yields of the signal and normalization modes. The index $i$ designates the fit category for the three $D^{*}$ decays. Only statistical uncertainties are given.
    }
    \begin{tabular}{llrlrlrl}
        \toprule
        \toprule
        Parameter & & \multicolumn{6}{c}{Observed (expected) yield} \\
        \midrule
        & & \multicolumn{2}{c}{${D^{*}}^{+} \rightarrow D^{0}\pi^{+}$} & \multicolumn{2}{c}{${D^{*}}^{+} \rightarrow D^{+}\pi^{0}$} & \multicolumn{2}{c}{${D^{*}}^{0} \rightarrow D^{0}\pi^{0}$} \\
        \cmidrule{3-8}
        $N_{D^{*}\tau\nu}^{i} + N_{D^{*}\tau\nu,\ell\text{-}\mathrm{misID}}^{i}$ & & \multicolumn{2}{c}{$50.9 \pm 7.8$} & \multicolumn{2}{c}{$7.8 \pm 1.2$} & \multicolumn{2}{c}{$49.2 \pm 7.5$} \rule[0mm]{0mm}{2.0mm} \\
        $N_{D^{*}\ell\nu}^{i}$ & & $1084.6 \pm 36.7$ & ($1041.0 \pm 11.2$)~~ & $137.9 \pm 6.6$ & ($133.2 \pm 4.3$)~~ & $940.9 \pm 36.0$ & ($927.2 \pm 10.7$)~~ \rule[0mm]{0mm}{4.0mm} \\
        \bottomrule
        \bottomrule
    \end{tabular}
    \label{tab:fittedYields}
\end{table*}

\section{Systematic Uncertainties}
\label{sec:sys}

The systematic uncertainties are listed in Table~\ref{tab:systUncert}. In general, for each source of uncertainty, we evaluate the shift $\Delta R(D^{*})$ of the $R(D^{*})$ result observed in data resulting from the relevant model variation. When the resulting $\Delta R(D^{*})$ distribution approximates a symmetric or an asymmetric Gaussian distribution, we use the observed standard deviation as systematic uncertainty. In other cases, we determine the systematic uncertainty as the maximum and minimum observed shifts in $\Delta R(D^{*})$.

The systematic uncertainty due to the PDF shapes is introduced to account for the uncertainty in the correction for the differences between data and simulation in the shapes of the $E_{\mathrm{ECL}}$ and $M_{\mathrm{miss}}^{2}$ distributions. The $E_{\mathrm{ECL}}$ energy shift is determined with an uncertainty ranging from $-7~\mathrm{MeV}$ to $+9~\mathrm{MeV}$ from the nominal shift. The smearing factors derived to correct for the $M_{\mathrm{miss}}^{2}$ resolution have statistical uncertainties of ${}_{-0.017}^{+0.015}$ $\left({}_{-0.054}^{+0.028}\right)~\mathrm{GeV}^{2}/c^{4}$ for neutral (charged) $B$ modes. These corrections modify the fit PDFs. To estimate the systematic uncertainty arising from variations in the PDF shapes, we generate PDFs with different energy shifts and smearing factors. The energy shift applied to the hadronic split-off showers is changed from $-7~\mathrm{MeV}$ to $+9~\mathrm{MeV}$ in $2~\mathrm{MeV}$ steps. Additionally, the smearing factors are varied 1000 times based on their statistical uncertainties at each energy shift. The resulting PDFs modified based on alternative values of the energy shifts and smearing factors are then used in alternative fits to data. The  ${}_{-8.3\%}^{+9.1\%}$ standard deviation of the distribution of the resulting $\Delta R(D^{*})$ values is assigned as the associated systematic uncertainty.

To estimate the systematic uncertainty from the limited size of the simulation sample, we employ the bootstrap sampling method~\cite{BS:1988}. For each PDF of a candidate category in every $D^{*}$ decay mode, we randomly resample with replacement the simulation sample. In the resampling, the number of events for a category follows a Poisson distribution. Furthermore, the efficiencies, $\varepsilon_{D^{*}\tau\nu}$ and $\varepsilon_{D^{*}\ell\nu}$, are updated for each fit. Subsequently, alternative PDFs are constructed based on the bootstrap samples and used to fit to the data. The value of $\Delta R({D^{*}})$ is obtained from the difference between $R({D^{*}})$ results obtained in the bootstrap samples and in the original simulation sample. The procedure is repeated 1000 times and the $7.5\%$ standard deviation of the $\Delta R({D^{*}})$ distribution is assigned as a systematic uncertainty. 

The systematic uncertainty due to uncertainties in the composition of $\overline{B} \rightarrow D^{**}\ell^{-}\overline{\nu}_{\ell}$ decays is determined by varying the measured branching fractions of the resonant $D^{**}\overline{\nu}_{\ell}$ and nonresonant $D^{*}\pi(\pi)\ell^{-}\overline{\nu}_{\ell}$ and $D_{s}^{(*)}K\ell^{-}\overline{\nu}_{\ell}$ decays based on their total uncertainties. If a branching fraction becomes negative in the variation, we exclude its decay mode. The unmeasured branching fractions of gap modes $B \rightarrow D_{\mathrm{gap}}^{**}\ell^{-}\overline{\nu}_{\ell}$ and $\overline{B} \rightarrow D^{**}\tau^{-}\overline{\nu}_{\tau}$ are varied uniformly from 0\% to 200\% of their estimated branching fractions. The maximum and minimum shifts in $R(D^{*})$ are taken as the systematic uncertainties for nonresonant $D^{*}\pi(\pi)\ell^{-}\overline{\nu}_{\ell}$ and the unmeasured categories, while the standard deviations of the $\Delta R(D^{*})$ distributions are taken for the other measured categories. The quadratic sum of the uncertainties due to the measured and unmeasured decay modes, ${}_{-3.5\%}^{+4.8\%}$, is assigned as the uncertainty from $\overline{B} \rightarrow D^{**}\ell^{-}\overline{\nu}_{\ell}$ branching fractions. 

The backgrounds from ${B}^{0}\leftrightarrow B^{+}$ cross feed, continuum, and ``other'' background categories are accompanied by an incorrectly reconstructed $B_{\mathrm{tag}}$ candidate. To account for possible discrepancies between data and simulation in the fraction of incorrectly reconstructed $B_{\mathrm{tag}}$ candidates, we vary the fixed yields of these categories across all $D^{*}$ modes from $0\%$ to $200\%$ of the expected yields in the simulation. The variation is repeated 1000 times and the maximum and minimum shifts observed in $\Delta R(D^{*})$ are assigned as the systematic uncertainty for each of the background categories. These uncertainties are combined in a quadratic sum for all three categories, resulting in a total systematic uncertainty of ${}_{-2.3\%}^{+2.7\%}$.

A similar procedure is employed to determine the uncertainty from the composition of the hadronic $B$ decay background. The branching fractions of $\overline{B} \rightarrow D^{*}\overline{D}{}_{s}^{(*)}$ and $\overline{B} \rightarrow D^{*}\mathrm{n}\pi(\pi^{0})$ decays are varied by their uncertainties according to a single Gaussian distribution to obtain $\Delta R(D^{*})$. Uncertainties between $\overline{B} \rightarrow D^{*}\overline{D}{}_{s}^{(*)}$ decays are assumed to be fully correlated while those between $\overline{B} \rightarrow D^{*}\mathrm{n}\pi(\pi^{0})$ decays are treated as uncorrelated. The correlation for $\overline{B} \rightarrow D^{*}\overline{D}{}_{s}^{(*)}$ decays takes into account the systematic variation due to cross feed in the branching fraction measurement~\cite{BaBarDsDBF:2007}. Contributions of hadronic $B$ decays that are not measured are also varied from $0\%$ to $200\%$ of their estimated branching fraction, while $\overline{B} \rightarrow D^{*}\overline{D}{}^{(*)}K$ decays are not considered because they contribute only a small fraction to the total background. The total uncertainty from all hadronic $B$ decays is $2.1\%$, which is the quadratic sum of the individual sources.

Systematic uncertainties arise from various efficiency corrections applied to the signal and normalization channels. These include the correction of the FEI reconstruction efficiency and the efficiency corrections due to track reconstruction, lepton and hadron identification, as well as the low-momentum $\pi$, $K_{S}^{0}$, and  $\pi^{0}$ reconstruction. Each of the efficiency corrections is varied by $\pm 1\sigma$ and the resulting differences in the PDF shapes are determined. The systematic uncertainty is $2.0\%$, which is obtained by adding these differences in quadrature. 

The KDE smooths the template histograms using a user-specified width scale factor for local densities. The PDF shape depends on the assigned value of this scale factor. To determine the systematic uncertainty associated with the KDE, the PDFs after the KDE are fit to simplified simulated experiments where KDE is not applied. The fit is repeated for 1000 simplified simulated experiments, and the observed shift in the $\Delta R(D^{*})$ distribution is taken as the systematic uncertainty of ${}_{-0.8\%}^{+2.0\%}$.

The form factors for the semileptonic $B$ decay models used in the simulation impact the distributions of kinematic quantities, such as $q^{2}$, and thus the PDF shapes in the final fit. To determine the associated systematic uncertainty, the $1\sigma$ uncertainties on the weights used for the form factor weighting are employed to construct covariance matrices for each signal $D^{*}$ decay and each category of semileptonic $B$ decays. An alternative PDF is then constructed by random sampling from the resulting covariance matrices. The varied PDF is used in an alternative $R(D^{*})$ fit to determine $\Delta R({D^{*}})$. A systematic uncertainty of ${}_{-0.1\%}^{+0.5\%}$ is assigned.

There is a small peaking component in the $\Delta M_{D^{*}}$ distribution of the fake $D^{*}$ candidate contribution. More than 90\% of this peaking component comes from incorrectly reconstructed $\overline{B} \rightarrow D^{*}\ell^-\overline{\nu}_{\ell}$ events.  The main sources of misreconstruction are incorrect assignment of the charged low-momentum pion and $D$ meson misreconstruction due to the inclusion of photon candidates from beam-induced background or hadronic split-off showers. The first source is expected to cancel in the $R(D^{*})$ ratio. The second source may not be well modeled by simulation and thus results in a systematic uncertainty. We vary the normalization of the peaking background contribution, where $\pi^{0}$ daughters of the $D$ meson are misreconstructed, from 0\% to 200\%, and assign the resulting shift in $R(D^{*})$, 0.4\%, as the corresponding systematic uncertainty.

The uncertainties in the branching fractions of leptonic $\tau$ decays can induce changes in $R(D^{*})$ due to variations in signal efficiency. We repeatedly fluctuate the branching fractions 1000 times, using a Gaussian function with a standard deviation equal to their known uncertainties~\cite{pdg:2022}. The standard deviation of the resulting $\Delta R({D^{*}})$ distributions, amounting to $0.2\%$, is assigned as a systematic uncertainty.

Finally, we account for the systematic uncertainties induced by the $R(D^{*})$ fit. The fit bias is $-0.1\%$ at $R(D^{*}) = 0.262$ using the linearity function of Eq.~\ref{eq:fitterLinearity}. Furthermore, there is a discrepancy observed between data and simulation in the range $1.8 < E_{\mathrm{ECL}} < 2.0~\mathrm{GeV}$. When this range is excluded, the $p$-value for the goodness of fit used in the $R(D^{*})$ extraction increases from $4.4\%$ to $14.4\%$. Reducing the fit range results in a $+0.1\%$ shift of the fitted $R(D^{*})$. The systematic uncertainty is determined by a quadratic sum of these two contributions, yielding ${}_{-0.1\%}^{+0.1\%}$.

\begin{table}[h]
    \centering
    \caption{Summary of systematic uncertainties on $R(D^{*})$. }
    \begin{tabular}{lr}
        \toprule
        \toprule
        Source & Uncertainty \\
        \midrule
        PDF shapes & ${}_{-8.3\%}^{+9.1\%}$ \rule[0mm]{0mm}{2mm} \\
        Simulation sample size & ${}_{-7.5\%}^{+7.5\%}$ \rule[0mm]{0mm}{4mm}\\
        $\overline{B} \rightarrow D^{**}\ell^{-}\overline{\nu}_{\ell}$ branching fractions & ${}_{-3.5\%}^{+4.8\%}$ \rule[0mm]{0mm}{4mm} \\
        Fixed backgrounds & ${}_{-2.3\%}^{+2.7\%}$ \rule[0mm]{0mm}{4mm} \\
        Hadronic $B$ decay branching fractions & ${}_{-2.1\%}^{+2.1\%}$ \rule[0mm]{0mm}{4mm} \\
        Reconstruction efficiency & ${}_{-2.0\%}^{+2.0\%}$ \rule[0mm]{0mm}{4mm} \\
        Kernel density estimation & ${}_{-0.8\%}^{+2.0\%}$ \rule[0mm]{0mm}{4mm} \\
        Form factors & ${}_{-0.1\%}^{+0.5\%}$ \rule[0mm]{0mm}{4mm} \\
        Peaking background in $\Delta M_{D^{*}}$ & ${}_{-0.4\%}^{+0.4\%}$ \rule[0mm]{0mm}{4mm} \\
        $\tau^{-} \rightarrow \ell^{-}\nu_{\tau}\bar{\nu}_{\ell}$ branching fractions & ${}_{-0.2\%}^{+0.2\%}$ \rule[0mm]{0mm}{4mm} \\
        $R(D^{*})$ fit method & ${}_{-0.1\%}^{+0.1\%}$ \rule[0mm]{0mm}{4mm} \\
        \midrule 
        Total systematic uncertainty & ${}_{-12.3\%}^{+13.5\%}$ \rule[0mm]{0mm}{2mm} \\
        \bottomrule
        \bottomrule
    \end{tabular}
    \label{tab:systUncert}
\end{table}

\section{Discussion and Conclusion}
\label{sec:conclusion}

We present a measurement of $R(D^{*}) = \mathcal{B}(\overline{B} \rightarrow D^{*} \tau^{-} \overline{\nu}_{\tau})$/$\mathcal{B} (\overline{B} \rightarrow D^{*} \ell^{-} \overline{\nu}_{\ell})$ using $189~\rm{fb}^{-1}$ of electron-positron collision data recorded at the $\Upsilon(\mathrm{4S})$ resonance by the Belle~II detector. A tag $B$ meson is fully reconstructed in a hadronic decay, and and the partner signal decay is reconstructed as $\overline{B}\rightarrow D^{*} \tau^{-}\overline{\nu}_{\tau}$ using leptonic $\tau$ decays. We find
\begin{equation}
R(D^{*}) = 0.262~_{-0.039}^{+0.041}(\mathrm{stat})~_{-0.032}^{+0.035}(\mathrm{syst}),
\end{equation}
where the first uncertainty is statistical and the second uncertainty is systematic. 
This is the first $R(D^{*})$ measurement from the Belle~II experiment. The statistical uncertainty of this measurement, $_{-14.7\%}^{+15.7\%}$, is comparable in precision to the corresponding Belle result (13.0\%)~\cite{BelleDtaunu1}, despite being based on a much smaller data sample ($189~\mathrm{fb}^{-1}$ compared to $711~\rm{fb}^{-1}$). This improved sensitivity is due to the use of a new $B$ tagging algorithm and an optimized selection. The Belle~II $R(D^{*})$ result is consistent with the current world average of these measurements and with SM predictions~\cite{HFLAV}, \footnote{Including the Belle~II $R(D^{*})$ result, the new world averages of all measurements differ by $2.2\sigma$ and $3.3\sigma$ from the SM predictions for $R(D^{*})$ and ($R(D)$, $R(D^{*})$).}.

\begin{acknowledgments}
This work, based on data collected using the Belle II detector, which was built and commissioned prior to March 2019, was supported by
Higher Education and Science Committee of the Republic of Armenia Grant No.~23LCG-1C011;
Australian Research Council and Research Grants
No.~DP200101792, 
No.~DP210101900, 
No.~DP210102831, 
No.~DE220100462, 
No.~LE210100098, 
and
No.~LE230100085; 
Austrian Federal Ministry of Education, Science and Research,
Austrian Science Fund
No.~P~31361-N36
and
No.~J4625-N,
and
Horizon 2020 ERC Starting Grant No.~947006 ``InterLeptons'';
Natural Sciences and Engineering Research Council of Canada, Compute Canada and CANARIE;
National Key R\&D Program of China under Contract No.~2022YFA1601903,
National Natural Science Foundation of China and Research Grants
No.~11575017,
No.~11761141009,
No.~11705209,
No.~11975076,
No.~12135005,
No.~12150004,
No.~12161141008,
and
No.~12175041,
and Shandong Provincial Natural Science Foundation Project~ZR2022JQ02;
the Czech Science Foundation Grant No.~22-18469S;
European Research Council, Seventh Framework PIEF-GA-2013-622527,
Horizon 2020 ERC-Advanced Grants No.~267104 and No.~884719,
Horizon 2020 ERC-Consolidator Grant No.~819127,
Horizon 2020 Marie Sklodowska-Curie Grant Agreement No.~700525 ``NIOBE''
and
No.~101026516,
and
Horizon 2020 Marie Sklodowska-Curie RISE project JENNIFER2 Grant Agreement No.~822070 (European grants);
L'Institut National de Physique Nucl\'{e}aire et de Physique des Particules (IN2P3) du CNRS
and
L'Agence Nationale de la Recherche (ANR) under grant ANR-21-CE31-0009 (France);
BMBF, DFG, HGF, MPG, and AvH Foundation (Germany);
Department of Atomic Energy under Project Identification No.~RTI 4002,
Department of Science and Technology,
and
UPES SEED funding programs
No.~UPES/R\&D-SEED-INFRA/17052023/01 and
No.~UPES/R\&D-SOE/20062022/06 (India);
Israel Science Foundation Grant No.~2476/17,
U.S.-Israel Binational Science Foundation Grant No.~2016113, and
Israel Ministry of Science Grant No.~3-16543;
Istituto Nazionale di Fisica Nucleare and the Research Grants BELLE2;
Japan Society for the Promotion of Science, Grant-in-Aid for Scientific Research Grants
No.~16H03968,
No.~16H03993,
No.~16H06492,
No.~16K05323,
No.~17H01133,
No.~17H05405,
No.~18K03621,
No.~18H03710,
No.~18H05226,
No.~19H00682, 
No.~20H05850,
No.~20H05858,
No.~22H00144,
No.~22K14056,
No.~22K21347,
No.~23H05433,
No.~26220706,
and
No.~26400255,
the National Institute of Informatics, and Science Information NETwork 5 (SINET5), 
and
the Ministry of Education, Culture, Sports, Science, and Technology (MEXT) of Japan;  
National Research Foundation (NRF) of Korea Grants
No.~2016R1\-D1A1B\-02012900,
No.~2018R1\-A2B\-3003643,
No.~2018R1\-A6A1A\-06024970,
No.~2019R1\-I1A3A\-01058933,
No.~2021R1\-A6A1A\-03043957,
No.~2021R1\-F1A\-1060423,
No.~2021R1\-F1A\-1064008,
No.~2022R1\-A2C\-1003993,
and
No.~RS-2022-00197659,
Radiation Science Research Institute,
Foreign Large-Size Research Facility Application Supporting project,
the Global Science Experimental Data Hub Center of the Korea Institute of Science and Technology Information
and
KREONET/GLORIAD;
Universiti Malaya RU grant, Akademi Sains Malaysia, and Ministry of Education Malaysia;
Frontiers of Science Program Contracts
No.~FOINS-296,
No.~CB-221329,
No.~CB-236394,
No.~CB-254409,
and
No.~CB-180023, and SEP-CINVESTAV Research Grant No.~237 (Mexico);
the Polish Ministry of Science and Higher Education and the National Science Center;
the Ministry of Science and Higher Education of the Russian Federation
and
the HSE University Basic Research Program, Moscow;
University of Tabuk Research Grants
No.~S-0256-1438 and No.~S-0280-1439 (Saudi Arabia);
Slovenian Research Agency and Research Grants
No.~J1-9124
and
No.~P1-0135;
Agencia Estatal de Investigacion, Spain
Grant No.~RYC2020-029875-I
and
Generalitat Valenciana, Spain
Grant No.~CIDEGENT/2018/020;
National Science and Technology Council,
and
Ministry of Education (Taiwan);
Thailand Center of Excellence in Physics;
TUBITAK ULAKBIM (Turkey);
National Research Foundation of Ukraine, Project No.~2020.02/0257,
and
Ministry of Education and Science of Ukraine;
the U.S. National Science Foundation and Research Grants
No.~PHY-1913789 
and
No.~PHY-2111604, 
and the U.S. Department of Energy and Research Awards
No.~DE-AC06-76RLO1830, 
No.~DE-SC0007983, 
No.~DE-SC0009824, 
No.~DE-SC0009973, 
No.~DE-SC0010007, 
No.~DE-SC0010073, 
No.~DE-SC0010118, 
No.~DE-SC0010504, 
No.~DE-SC0011784, 
No.~DE-SC0012704, 
No.~DE-SC0019230, 
No.~DE-SC0021274, 
No.~DE-SC0021616, 
No.~DE-SC0022350, 
No.~DE-SC0023470; 
and
the Vietnam Academy of Science and Technology (VAST) under Grants
No.~NVCC.05.12/22-23
and
No.~DL0000.02/24-25.

These acknowledgements are not to be interpreted as an endorsement of any statement made
by any of our institutes, funding agencies, governments, or their representatives.

We thank the SuperKEKB team for delivering high-luminosity collisions;
the KEK cryogenics group for the efficient operation of the detector solenoid magnet;
the KEK computer group and the NII for on-site computing support and SINET6 network support;
and the raw-data centers at BNL, DESY, GridKa, IN2P3, INFN, and the University of Victoria for off-site computing support.
\end{acknowledgments}

\bibliographystyle{myapsrev4-2}
\bibliography{references}  

\end{document}